\documentclass{aa}  

\usepackage{graphicx}
\usepackage{subcaption}
\usepackage{txfonts}
\usepackage{hyperref}

\usepackage{suffix}
\usepackage[printonlyused]{acro}

\acsetup{
  single        = false,
  list/sort     = true,
  cite/group    = true,
  cite/group/cmd = \citealp,
  cite/group/pre = {; },
}

\DeclareAcronym{EVE}{
  short = EVE,
  long = \textit{Extreme Ultraviolet Variability Experiment},
  cite = {Woods:2012}}

\DeclareAcronym{AIA}{
  short = AIA,
  long = \textit{Atmospheric Imaging Assembly},
  cite = {Lemen:2012}}

\DeclareAcronym{HMI}{
  short = HMI,
  long = \textit{Helioseismic Magnetic Imager},
  cite = {Scherrer:2012}}
  
\DeclareAcronym{SDO}{
  short = SDO,
  long = \textit{Solar Dynamics Observatory},
  cite = {Pesnell:2012}}

\DeclareAcronym{STEREO}{
  short = STEREO,
  long = \textit{Solar Terrestrial Relations Observatory},
  cite = {Kaiser:2008}}

\DeclareAcronym{SECCHI}{
  short = SECCHI,
  long = \textit{Sun Earth Connection Coronal and Heliospheric Investigation},
  cite = {Howard:2008}}
  
\DeclareAcronym{SCIP}{
  short = SCIP,
  long = Sun Centered Imaging Package}
  
\DeclareAcronym{MGN}{
  short = MGN,
  long = multi-scale Gaussian normalisation,
  cite = {Morgan:2014}}

\DeclareAcronym{EUVI}{
  short = EUVI,
  long = EUVI}

\DeclareAcronym{CME}{
  short = CME,
  short-plural-form = CMEs,
  long = coronal mass ejection,
  long-plural-form = coronal mass ejections}
  
\DeclareAcronym{PIL}{
  short = PIL,
  short-plural-form = PILs,
  long = polarity inversion line,
  long-plural-form = polarity inversion lines}
  
\DeclareAcronym{LOS}{
  short = LOS,
  short-plural-form = LOS,
  long = line of sight,
  long-plural-form = lines of sight}
  
\DeclareAcronym{FOV}{
  short = FOV,
  short-plural-form = FOV,
  long = field of view,
  long-plural-form = fields of view}
  
\DeclareAcronym{UV}{
  short = UV,
  short-plural-form = UV,
  long = ultraviolet,
  long-plural-form = ultraviolet}

\DeclareAcronym{EUV}{
  short = EUV,
  short-plural-form = EUV,
  long = extreme ultraviolet,
  long-plural-form = extreme ultraviolet}
  
\DeclareAcronym{GONG}{
  short = GONG,
  long = \textit{Global Oscillations Network Group}}
  
\DeclareAcronym{DST}{
  short = DST,
  long = \textit{Dunn Solar Telescope}}

\DeclareAcronym{IBIS}{
  short = IBIS,
  long = \textit{Interferometric Bidimensional Spectropolarimeter},
  cite = {Cavallini:2006,Reardon:2008}}

\DeclareAcronym{FIRS}{
  short = FIRS,
  long = \textit{Facility Infrared Spectropolarimeter},
  cite = {Jaeggli:2011}}
  
\DeclareAcronym{EIT}{
  short = EIT,
  long = \textit{Extreme Ultraviolet Imaging Telescope},
  cite = {Delaboudiniere:1995}}
 
\DeclareAcronym{ROSA}{
  short = ROSA,
  long = \textit{Rapid Oscillations in the Solar Atmosphere},
  cite = {Jess:2010}}
  
\DeclareAcronym{SPINOR}{
  short = SPINOR,
  long = \textit{Spectro-Polarimeter for Infrared and Optical Regions},
  cite = {Socas:2006}}  
 
\DeclareAcronym{SOHO}{
  short = SOHO,
  long = \textit{Solar and Heliospheric Observatory},
  cite = {Domingo:1995}}
 
\DeclareAcronym{MDI}{
  short = MDI,
  long = \textit{Michelson Doppler Imager},
  cite = {Scherrer:1995}}

\DeclareAcronym{NMSU}{
  short = NMSU,
  long = New Mexico State University}

\DeclareAcronym{NSO}{
  short = NSO,
  long = \textit{National Solar Observatory}}
  
\DeclareAcronym{TAC}{
  short = TAC,
  long = time allocation committee}
  
\DeclareAcronym{DEM}{
  short = DEM,
  short-plural-form = DEMs,
  long = Differential Emission Measure,
  long-plural-form = Differential Emission Measures}
  
\DeclareAcronym{PI}{
  short = PI,
  short-plural-form = PI,
  long = Principle Investigator,
  long-plural-form = Principle Investigators}  
  
\DeclareAcronym{IDL}{
  short = IDL,
  long = Interactive Data Language}
  
\DeclareAcronym{bb}{
  short = bb,
  long = broadband}  
  
\DeclareAcronym{nb}{
  short = nb,
  long = narrowband}    
  
\DeclareAcronym{IPM}{
  short = IPM,
  long = interplanetary medium}  
  
\DeclareAcronym{AO}{
  short = AO,
  long = adaptive optics,
  cite = {Rimmele:2004}} 
  
\DeclareAcronym{IR}{
  short = IR,
  long = infrared}  

\DeclareAcronym{halpha}{
  short = H-$\alpha$,
  long = Hydrogen-$\alpha$}

\DeclareAcronym{FPI}{
  short = FPI,
  long = Fabry-P\'erot}
  
\DeclareAcronym{DWDM}{
  short = DWDM,
  long = dense wavelength division multiplexing}  

\DeclareAcronym{CCD}{
  short = CCD,
  long = charge-coupled device}   
  
\DeclareAcronym{HI}{
  short = HI,
  long = Heliospheric Investigation}   
  
\DeclareAcronym{ROI}{
  short = ROI,
  long = region-of-interest}    
  
\DeclareAcronym{POV}{
  short = POV,
  long = point-of-view}  
  
\DeclareAcronym{MHS}{
  short = MHS,
  long = magnetohydrostatic}
  
\DeclareAcronym{MHD}{
  short = MHD,
  long = magnetohydrodynamic}  
  
\DeclareAcronym{RMHD}{
  short = RMHD,
  long = radiative magnetohydrodynamic}  
  
\DeclareAcronym{DOT}{
  short = DOT,
  long = \textit{Dutch Open Telescope},
  cite = {Rutten:1997}}  
  
\DeclareAcronym{BBSO}{
  short = BBSO,
  long = \textit{Big Bear Solar Observatory}}
  
\DeclareAcronym{NST}{
  short = NST,
  long = \textit{New Solar Telescope},
  cite = {Goode:2012}}

\DeclareAcronym{SOT}{
  short = SOT,
  long = \textit{Solar Optical Telescope},
  cite = {Tsuneta:2008}}
  
\DeclareAcronym{hinode}{
  short = Hinode,
  long = Hinode,
  cite = {Kosugi:2007}}
  
\DeclareAcronym{GST}{
  short = GST,
  long = \textit{Goode Solar Telescope}}  
  
\DeclareAcronym{SST}{
  short = SST,
  long = \textit{Swedish 1-m Solar Telescope},
  cite = {Scharmer:2002}}  
  
\DeclareAcronym{TRACE}{
  short = TRACE,
  long = \textit{Transition Region and Coronal Explorer},
  cite = {Handy:1999}}  
  
\DeclareAcronym{PCTR}{
  short = PCTR,
  long = \textit{prominence-corona-transition-region}}
  
\DeclareAcronym{DKIST}{
  short = DKIST,
  long = Daniel K. Inouye Solar Telescope}
  
\DeclareAcronym{RTI}{
  short = RTI,
  long = Rayleigh-Taylor instability}
  
\DeclareAcronym{SSW}{
  short = SSW,
  long = SolarSoftWare,
  cite = {Freeland:1998}}    
  
\DeclareAcronym{LTE}{
  short = LTE,
  long = local thermodynamic equilibrium}
  
\DeclareAcronym{NLTE}{
  short = NLTE,
  long = non-"local thermodynamic equilibrium"}
  
\DeclareAcronym{FTS}{
  short = FTS,
  long = Fourier Transform Spectrometer,
  cite = {Kurucz:1984}}
  
\DeclareAcronym{RTE}{
  short = RTE,
  long = radiative transfer equation}
  
\DeclareAcronym{CRD}{
  short = CRD,
  long = complete frequency redistribution}
  
\DeclareAcronym{PRD}{
  short = PRD,
  long = partial frequency redistribution}
  
\DeclareAcronym{BCM}{
	short = BCM,
	long = Beckers' cloud model,
	cite={Beckers:1964}}
	
\DeclareAcronym{HSRA}{
	short = HSRA,
	long = Harvard Smithsonian Reference Atmosphere,
	cite={Gingerich:1971}}
	
\DeclareAcronym{NICOLE}{
	short = NICOLE,
	long = Non-LTE Inversion COde using the Lorien Engine,
	cite={Socasnavarro:2015}}
	
\DeclareAcronym{SIR}{
	short = SIR,
	long = Stokes Inversion based on Response functions,
	cite={Ruizcobo:1992}}
	
\DeclareAcronym{HAZEL}{
	short = HAZEL,
	long = Hanle and Zeeman Light,
	cite={AsensioRamos:2008}}
	
\DeclareAcronym{BPSS}{
	short = BPSS,
	long = bald patch separatrix surface,
	cite={Bungey:1996}}
	
\DeclareAcronym{EIS}{
	short = EIS,
	long = \textit{EUV Imaging Spectrometer},
	cite={Culhane:2007}}

\DeclareAcronym{FWHM}{
  short = FWHM,
  long = full width at half maximum}

\DeclareAcronym{AMRVAC}{
	short = MPI-AMRVAC,
	long = \textit{Adaptive Mesh Refinement Versatile Advection Code},
	cite = {Keppens:2012,Porth:2014,Xia:2018,Keppens:2020}}

\DeclareAcronym{AMR}{
	short = AMR,
	long = adaptive mesh refinement}

\DeclareAcronym{CCI}{
	short = CCI,
	long = Convective Continuum Instability}

\DeclareAcronym{BV}{
	short = BV,
	long = Brunt-V\"ais\"al\"a}

\DeclareAcronym{TVDLF}{
	short = TVDLF,
	long = Total Variation Diminishing Lax-Friedrich}

\DeclareAcronym{TI}{
	short = TI,
	long = Thermal Instability}

\DeclareAcronym{TNE}{
	short = TNE,
	long = thermal non-equilibrium}
	
\DeclareAcronym{MALI}{
    short = MALI,
    long = Multilevel Accelerated Lambda Iteration}
    
\DeclareAcronym{yt}{
    short = yt,
    long = yt-project,
    cite = {Turk:2011}}
    
\DeclareAcronym{GL}{
    short = GL,
    long = Gauss-Legendre}
    
\DeclareAcronym{CM}{
    short = CM,
    long = classically mounted}
    
\DeclareAcronym{MULTI3D}{
    short = MULTI3D,
    long = \textit{multi-level non-LTE 3D},
    cite = {Leenaarts:2009}}

\DeclareAcronym{EB}{
    short = EB,
    long = \textit{Eddington-Barbier}}
    
\DeclareAcronym{KDE}{
    short = KDE,
    long = kernel density estimate}
    
\DeclareAcronym{NLFFF}{
    short = NLFFF,
    long = nonlinear force-free field}

\usepackage[normalem]{ulem}
\usepackage[x11names]{xcolor}
\usepackage{academicons}
\definecolor{orcidlogocol}{HTML}{A6CE39}

\newcommand{\orcid}[1]{\href{https://orcid.org/#1}{\textcolor[HTML]{A6CE39}{\aiOrcid}}}

\usepackage{xcolor}
\usepackage{soul}

\begin{document}

   \title{From eruption to post-flare rain: a 2.5D MHD model}

   \author{Samrat Sen$^*$ \inst{1, 2, 3} \href{https://orcid.org/0000-0003-1546-381X}{\includegraphics[scale=0.05]{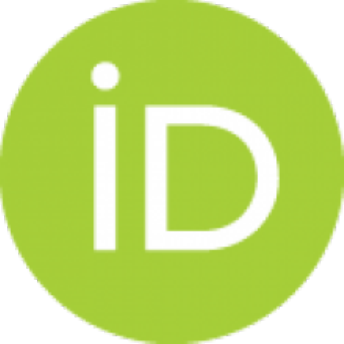}},  Avijeet Prasad \inst{4, 5}, Valeriia Liakh\inst{3}, Rony Keppens\inst{3} 
          }

   \institute{Instituto de Astrof\'{i}sica de Canarias, 38205 La Laguna, Tenerife, Spain
   \and
   Universidad de La Laguna, 38206 La Laguna, Tenerife, Spain
   \and
   Centre for mathematical Plasma-Astrophysics, Celestijnenlaan 200B, 3001 Leuven, KU Leuven, Belgium
   \and
    Rosseland Centre for Solar Physics, University of Oslo, Postboks 1029 Blindern, 0315 Oslo, Norway
    \and 
    Institute of Theoretical Astrophysics, University of Oslo, Postboks 1029 Blindern, 0315 Oslo, Norway    
   \\
   $^*$\email{samrat.sen@iac.es; samratseniitmadras@gmail.com}
                      }

   \date{Received: XXXX; accepted: XXXX}
   \date{}
 
  \abstract
   {Erupting magnetic flux ropes (MFRs) play an important role in producing solar flares, whereas fine-scale, condensed coronal rain is often found in post-flare loops. However, the formation of the MFRs in the pre-flare stage, and how this leads to coronal rain in a post-eruption magnetic loop is not fully understood.}
   {We explore the formation, and eruption of MFRs, followed by the appearance of coronal rain in the post-flare loops, to understand the magnetic and thermodynamic properties of eruptive events and their multi-thermal aspects in the solar atmosphere.}
   {We perform a resistive-magnetohydrodynamic (MHD) simulation with the open-source code \texttt{MPI-AMRVAC} to explore the evolution of sheared magnetic arcades that can lead to flux rope eruptions. The system is in mechanical imbalance at the initial state, and evolves self-consistently in a non-adiabatic atmosphere under the influence of radiative losses, thermal conduction, and background heating. We use an additional level of adaptive mesh refinement to achieve the smallest cell size of $\approx 32.7$ km in each direction to reveal the fine structures in the system.}
   {The system achieves a semi-equilibrium state after a short transient evolution from its initial mechanically imbalanced condition. A series of erupting MFRs is formed due to spontaneous magnetic reconnection, across current sheets created underneath the erupting flux ropes. Gradual development of thermal imbalance is noticed at a loop top in the post-eruption phase, which leads to catastrophic cooling and formation of condensations. We obtain plasma blobs which fall down along the magnetic loop in the form of coronal rain. The dynamical and thermodynamic properties of these cool-condensations are in good agreement with observations of post-flare coronal rain.}
   {The presented simulation supports the development and eruption of multiple MFRs, and the formation of coronal rain in post-flare loops, which is one of the key aspects to reveal the coronal heating mystery in the solar atmosphere.}
 

\titlerunning{Post-flare coronal rain}
\authorrunning{Sen et al.}

\maketitle


\section{Introduction} 

Eruptive events such as flares and coronal mass ejections (CMEs) occur due to the conversion of the stored magnetic energy of twisted magnetic field lines into the abrupt release of kinetic, thermal, and non-thermal energies \citep{1999:antiochos, 2020:fleishman}. Astrophysical plasmas, in general, have very high Lundquist numbers, $S_L = v_a L/\eta_m$ (where, $v_a$, $L$, and $\eta_m$ are the Alfv{\'e}n velocity, length scale, and magnetic diffusivity, respectively), and particularly for the solar corona, $S_L \sim 10^{10}$ \citep{2004:aschawnden-book}. Under such conditions,  Alfv{\'e}n's flux freezing theorem \citep{1942:alfven-fluxfreeze} is valid throughout most of the coronal volume, except in localized current sheets and in regions with strong magnetic gradients. Magnetic structures associated with the eruptive events can sometimes be described as magnetic flux ropes (MFRs), which are a bundle of twisted magnetic field lines winding about a common axis \citep{2011:chen, 2014:priest-book, 2022:he}. These MFRs can rise under favourable conditions through ideal magnetohydrodynamic (MHD) instabilities such as kink and torus instabilities \citep{2007:fan, 2006:kliem}, or through magnetic reconnection, where the magnetic field lines (MFLs) change their topology when they come close to each other in a resistive medium. The reconnection process converts the magnetic energy into thermal and kinetic energies and can generate acceleration of the charged particles below the flux ropes \citep{2003ApJ...585.1073A, 2012A&A...543A.110A}. A large fraction of this released energy can be transported downward from the corona to the denser chromosphere through thermal conduction and energetic electrons \citep{1978ApJ...220.1137A, 2020:wenzhi,2023:druett} in an arcade-like magnetic loop system. This energy deposition can lead to sudden heating of the local plasma in the chromosphere and causes upward evaporation into the corona to increase the density and temperature of the magnetic loop system \citep{2023:druett}. The loops gradually cool down to their usual coronal condition mainly through radiative cooling and field-aligned thermal conduction \citep{1995:cargill, 2001SoPh..204...91A}. These loops can be visible in extreme ultraviolet (EUV) \citep[and references therein]{2014:scullion, 2016:scullion, 2019:mason} and also in H$\alpha$ \citep{1964:bruzek} bands, and are called post-flare loops (PFLs).  

One of the most fascinating plasma processes in the solar corona is its multi-thermal behavior, where cool ($\sim 10^4$ K), and dense ($\sim 10^{10}$ cm$^{-3}$) materials appear in a background hot ($\sim 1$ MK) medium. This phenomenon mainly manifests itself in two distinct forms: prominence/filaments, and coronal rain, which are basically end results of the same underlying physics. Radiative processes are always at the origin of coronal rain or coronal condensation in general. When there is an abrupt cutoff in heating, which can happen after flares, the radiative cooling in a localized domain can dominate over heating. Temperature in that domain can drop down to $\sim 10^4$ K forming cool-condensed plasma that falls down towards the chromosphere in the form of rain \citep{1980:antiochos}. Coronal rain has been observed in PFLs in the classic ``loop prominence model" by \cite{1964:bruzek}. It is believed that condensations form in a PFL as a consequence of flare heating \citep{1980:antiochos}, and would lead to a single appearance and subsequent fall of coronal rain.
This is supported by the observations of \cite{2014:scullion, 2016:scullion} and \cite{2015:liu}. A related model studies loops undergoing ``thermal non-equilibrium" (TNE) as proposed by \cite{1991:antiochos} as the origin of long-lived prominence condensation. The idea is when the heating in the coronal loops is sufficiently localized at the loop foot points, the materials at the loop top cannot achieve thermal equilibrium. Then the loop materials can undergo a quasi-cyclic evolution of heating and chromospheric evaporation followed by catastrophic radiative cooling to form coronal rains. Active region loops can undergo this process to form many cycles of coronal rain, which is also supported by the observations of \cite{2015:patrick, 2015:froment} and \cite{2017:froment}. More recently, many 2D and 3D models have emerged \citep{2019A&A...624A..96C,2020A&A...636A.112C}, also of purely coronal volumes, where the TNE scenario is excluded by construction, and where the local trigger due to thermal instability is emphasized instead. Simulations of coronal rain in multidimensional arcade setups \citep{2015:fang, 2017A&A...603A..42X, 2022ApJ...926..216L} show a fair agreement with actual observations of rain in loops, and in essence combine the quasi-cyclic rain appearance with the thermal instability trigger process.

Towards realizing the aspect of coronal condensation after flux rope eruptions, we use a 2.5D resistive MHD simulation in a non-adiabatic description including optically thin radiative loss, (field-aligned) thermal conduction and steady (but spatially varying) background heating. To mimick coronal loop structures, we use a bipolar magnetic field configuration of sheared arcades, which are non-force-free and embedded in a stratified solar atmosphere with gravity. The Lorentz force (LF) of the system from the initial state drives a rapid evolution where one forms flux ropes from initial bipolar ﬁeld lines which ascend continuously towards eruption, mediated via the common process of spontaneous magnetic reconnections. The non-adiabatic effects of the system are crucial to trigger the gradual development of thermal instability, and the onset of the catastrophic cooling phase, which may lead to the formation of coronal condensation in general, and coronal rain in particular. 

Recent studies of reconnection-driven coronal condensations in solar current sheets in non-adiabatic regimes have been reported by \cite{2022A&A...666A..28S} in 2D geometry, and in a more complex topology of magnetic flux ropes in 3D geometry \citep{2023A&A...678A.132S}. Those studies did not include gravitational stratification, and highlighted the interplay between tearing and thermal instabilities in coronal current sheet setups. The formation of coronal rain in a post-eruption coronal loop has received less attention to date. A study by \cite{2021:wenzhi} shows the formation of coronal rain demonstrating multiple post-flare rain cycles in a post-flare loop. However, the eruption of MFRs is not captured in their study. In this work, we include the self-consistent development and eruption of a series of MFRs, followed by the formation of coronal rain in a loop top that gradually evolves with time, splits into different rain blobs, and falls downwards. We investigate the magnetic and thermodynamic properties of this event in detail to understand their formation mechanisms and the fine-scale structures to reveal the multi-thermal aspect of the solar atmosphere. 

The rest of the paper is organized as follows. In Section \ref{sec:setup}, we describe the numerical model with the precise initial condition, the governing equations with the numerical schemes used, and the boundary conditions. In Section \ref{sec:results}, we report the main results, and present a detailed analysis of the underlying physics. Section \ref{sec:summary-discussion} summarizes the key results of this study, discusses the applicability of the model to the solar atmosphere, makes comparison with existing observations, and addresses the novelty of the work. We conclude how this work can be useful in a broader aspect to understand the solar atmosphere in future studies.

\section{Numerical setup}\label{sec:setup}

The MHD processes in the solar atmosphere are extremely dynamic and non-linear. To explore the small-scale dynamics, magnetic, and thermodynamic properties of eruptions, and associated condensations in the solar corona, we construct a non-adiabatic and resistive-MHD simulation in a 2.5D cartesian geometry using MPI-parallelised Adaptive Mesh Refinement Versatile Advection code (MPI-AMRVAC) \footnote{Open source at: \href{http://amrvac.org}{http://amrvac.org/}} \citep{2012JCoPh.231..718K, 2014ApJS..214....4P, 2018ApJS..234...30X, keppens2021, keppens2023}. 

\subsection{Model setup and governing equations} \label{sub:model}

To construct a magnetic field topology similar to sheared coronal loops, we use an initial magnetic field configuration motivated by \cite{2014:priest-book} and \cite{2016:sanjay},
\begin{align}
    B_x &= B_0 \ \text{sin}(k_x(x-x_0)) \ \text{exp}(-yk_x/\sigma)\,, \label{eq:bx}\\ \label{eq:by}
    B_y &= B_0 \sigma \ \text{cos}(k_x(x-x_0)) \ \text{exp}(-yk_x/\sigma)\,,\\ \label{eq:bz}
    B_z &= B_0 \ \text{tan}(\gamma) \ \text{sin}(k_x(x-x_0)) \ \text{exp}(-yk_x/\sigma) \,,
\end{align}
where, the spatial domain of the simulation extends from $x= 0$ to $2 \pi$ Mm, along the horizontal direction, and $y=0$ to 25 Mm along the vertical direction, whereas the $z$-direction has translational symmetry. The parameter, $k_x=\frac{2\pi}{L_x}$ (where, $L_x=2\pi$ Mm is the horizontal span of the simulation domain), which gives $k_x=1$ Mm$^{-1}$. The arcade is centered at the numerical domain by setting $x_0 = \pi/2$ Mm. The shearing angle of the projected field lines at the $y=0$ plane with the $+x$ axis is determined by 
\begin{align} \label{eq:phi}
    \gamma = \text{tan}^{-1} (B_z/B_x)\,.  
\end{align}
Note that, for $\gamma=0$, the guide field $B_z$ vanishes, and the field lines are tangential to the constant $z$ plane which leads to no shearing. The associated Lorentz force is given by,
\begin{align} \label{eq:LF}
     {\bf F_L} =& \ (\sigma^2-\text{sec}^2\gamma)B_0^2k_x \text{sin}\big(k_x(x-x_0)\big)\ \text{exp}\bigg(-\frac{2k_x y}{\sigma}\bigg)\\ \nonumber 
     & \times \bigg(\text{cos}\big(k_x(x-x_0)\big) \ {\bf \hat{x}} - \frac{1}{\sigma} \text{sin}\big(k_x(x-x_0)\big) \ {\bf \hat{y}}\bigg),
\end{align}
where, ${\bf \hat{x}}$, and ${\bf \hat{y}}$ are the unit vectors along $x$, and $y$ directions respectively. The advantage of choosing this magnetic field configuration is that the system becomes force-free for $\displaystyle{|\sigma| = \text{sec}(\gamma)}$, and non-force-free otherwise. The initial magnetic field configuration (Equations (\ref{eq:bx})-(\ref{eq:bz})) obeys the solenoidal condition, $\displaystyle{\nabla \cdot {\bf B} = 0}$. In order to keep this condition throughout the simulation, we use the parabolic diffusion method \citep{keppens2003, keppens2023}. To achieve a sufficient shearing in the field lines at the initial stage, we set the value of $\gamma = 72^{\circ}.5$. Furthermore, we set $\sigma =10$, which ensures the presence of non-zero Lorentz force at the initial state of the system. The parameter, $B_0$, decides the strength of the magnetic field, where the maximum field strength (at $y=0$), $B_{max} = B_0 \sigma$. To set the magnetic field strength similar to typical coronal loops, we use $B_0 = 9$ G, that gives $B_{max}=90$ G, which is comparable with observation by \cite{2021:brooks}, where they have reported a field strength in active region loops in the range of $60-150$ G. To obtain a typical solar coronal condition, we set the unit density, $\bar{\rho} = 2.34 \times 10^{-15}$ g cm$^{-3}$, and unit temperature, $\bar{T} = 10^6$ K.

We use as base resolution of the simulation 96 $\times$ 384 along the $x$ and $y$ directions respectively, with only one additional adaptive mesh refinement (AMR) level. This achieves the smallest cell size of $\approx$32.7 km in each direction. The (de-)refinement is based on the errors estimated by the gradients of the instantaneous density and of magnetic field components at each time step \citep{1987:Lohner}. We follow the time evolution of the system up to 83.37 minutes.
 
To explore the evolution of the system, the following MHD equations are solved numerically,
\begin{align} \label{eq:mhd1}
   & \frac{\partial \rho}{\partial t} + \nabla \cdot (\rho \, {\bf v}) = 0, \\ \label{eq:mhd2}
   & \frac{\partial (\rho \, {\bf v})}{\partial t}+ \nabla \cdot (\rho\, {\bf v}\otimes{\bf v}+p_{tot}\,{\bf I}-{\bf B}\otimes {\bf B})= \rho\, {\bf g},\\ \label{eq:mhd3}
   & \frac{\partial \mathcal{E}}{\partial t} + \nabla \cdot \left[\mathcal{E}\,{\bf v}+p_{tot}\,{\bf v}-({\bf B}\cdot {\bf v})\,{\bf B}\right]= \rho\, {\bf g}\cdot {\bf v} + \eta \, \textbf{J}^2 - {\bf B} \cdot \left[\nabla \times (\eta {\bf J})\right]\\\nonumber 
   & \hspace*{2cm} - \rho^2 \Lambda(T)+H_{bgr}+ \nabla \cdot \left[\kappa_{||}\,{\bf b}\, ({\bf b} \cdot \nabla)\, T\right],\\ \label{eq:mhd4}
   & \frac{\partial {\bf B}}{\partial t} + \nabla \cdot ({\bf v}\otimes {\bf B}-{\bf B}\otimes {\bf v}) + \nabla \times (\eta \,{\bf J}) = \mathbf{0}\,,\\ \label{eq:mhd5}
   & \nabla \cdot \textbf{B}=0 \,,\\ \label{eq:mhd6}
   & \textbf{J}=\nabla \times \textbf{B} \,.
\end{align}
Here, we use magnetic units with unit magnetic permeability, {\bf I} is the unit tensor, and $\rho$, $T$, {\bf v}, $\eta$ and ${\bf b}$ represent plasma density, temperature, velocity vector, resistivity, and unit magnetic field vector, respectively, and $\otimes$ is the symbol for the dyadic product of two vectors. A uniform resistivity of $\eta = 5 \times 10^{-5}$ (in code unit), which corresponds to $5.82 \times 10^{10}$ cm$^2$ s$^{-1}$ is used throughout the simulation domain. The associated Lundquist number, $S_L=Lv_a/\eta$ (where, $L$ is the length scale and $v_a$ is the Alfv{\'e}nic speed) $\sim 10^5$. We adopt the Spitzer-type thermal conductivity, purely aligned along the magnetic fields, ${\bf \kappa}_{||} = 10^{-6}\, T^{5/2}$ erg cm$^{-1}$ s$^{-1}$ K$^{-1}$. The total pressure, $p_{tot}$ is the sum of the plasma and magnetic pressure,
\begin{align}
    p_{tot} = p+\frac{B^2}{2},
\end{align}
where $p$ is the plasma pressure determined by the thermodynamic quantities through the ideal gas law of the system. The total energy density per unit volume, ${\mathcal{E}}$, is obtained as the sum of the internal, kinetic and magnetic energy densities, 
\begin{align}\label{eq:energy_density}
    \mathcal{E}=\frac{p}{\gamma -1} + \frac{\rho v^2}{2} + \frac{B^2}{2},
\end{align}
where $\gamma = 5/3$ is the ratio of the specific heats. 

The density along the vertical domain is stratified according to hydrostatic equilibrium with gravity where ${\bf g}=-g(y) \mathbf{e}_y$ and,
\begin{align}
    g(y) = g_0 \frac{R_{\odot}^2}{(R_{\odot}+y)^2},
\end{align}
where, $g_0 = 274$ m s$^{-2}$ is the gravitational acceleration at the solar surface, and $R_{\odot} = 695.7$ Mm is the radius of the Sun. We assume an isothermal atmosphere, $T_0 = 1$ MK at the initial state of the system. The isothermal condition is relevant as the vertical extension of the domain is $\approx 25$ Mm. The initial density variation follows from
\begin{align}\label{eq:HS_equlb}
    \rho_i(y) = \rho_0\ \text{exp}\left[-y/H(y)\right],
\end{align}
where, $\displaystyle{H(y) = \frac{\mathcal{R} T_0}{\mu g_0} \frac{R_{\odot} +y}{R_{\odot}}}$ is the density scale height, which is based on the gas constant $\mathcal{R}$, and mean molecular mass, $\mu$. The scale height at the base ($y=0$) of our simulation domain is $\approx 50$ Mm, which increases only up to $\approx 51.8$ Mm at the top boundary ($y=25$ Mm). To use a typical coronal condition, we set $\rho_0 = 3.2 \times 10^{-15}$ g cm$^{-3}$. 

To achieve a coronal condensation phase through thermal runaway process, we incorporate the non-adiabatic effects of optically thin radiation (fourth term at the RHS of equation (\ref{eq:mhd3})), background heating (fifth term at the RHS of equation (\ref{eq:mhd3})), and field-aligned thermal conduction (last term at the RHS of equation (\ref{eq:mhd3})). To compensate for the initial radiative cooling exactly with the background heating ($H_{bgr}$), we use 
\begin{align}\label{eq:Hbgr}
    H_{bgr} = \rho_i^2(y) \Lambda(T_0)\,, 
\end{align}
where, $\Lambda(T)$ is estimated from the cooling function model developed by \cite{2008ApJ...689..585C} and \cite{1972ARA&A..10..375D}. This initial choice ensures thermal balance between radiative cooling and background heat at the start, but the deviations will occur as this height-dependent heating is kept constant throughout the simulation. Thermal conduction is zero at the initial state of the system due to our isothermal atmosphere assumption, but plays a role in the posterior evolution away from isothermal conditions.    

\subsection{Numerical schemes and Boundary conditions} \label{sub:bcs}

The equations (\ref{eq:mhd1}-\ref{eq:mhd6}) are solved numerically using a three-step, third-order Runge-Kutta time integration method with the Van Leer flux limiter \cite{1974:vanLeer} and Total Variation Diminishing Lax-Friedrichs (TVDLF) flux scheme. We use the `splitting' of the magnetic field variable, ${\bf B}$, where it is decomposed into a steady background term and a time-dependent deviation part \citep{2018ApJS..234...30X}. 

We use periodic boundary conditions at the side boundaries. Magnetic fields are extrapolated at the top and bottom boundaries by third-order and second-order zero-gradient methods respectively. The pressure and density values at the bottom boundary are fixed according to their local initial values, as prescribed in \cite{2021A&A...646A.134J}. At the top boundary, they are set according to the hydrostatic assumption, as prescribed in \cite{zhao2017}. The velocity components are set anti-symmetric in the form of mirror reflections at the bottom boundary to fill the ghost cells. At the top boundary, we use third order extrapolation with zero-gradient condition for all the velocity components, additionally with no inflows.

\begin{figure}[hbt!]
    \centering
\begin{subfigure}{0.92\columnwidth}
  \includegraphics[width=1\textwidth]{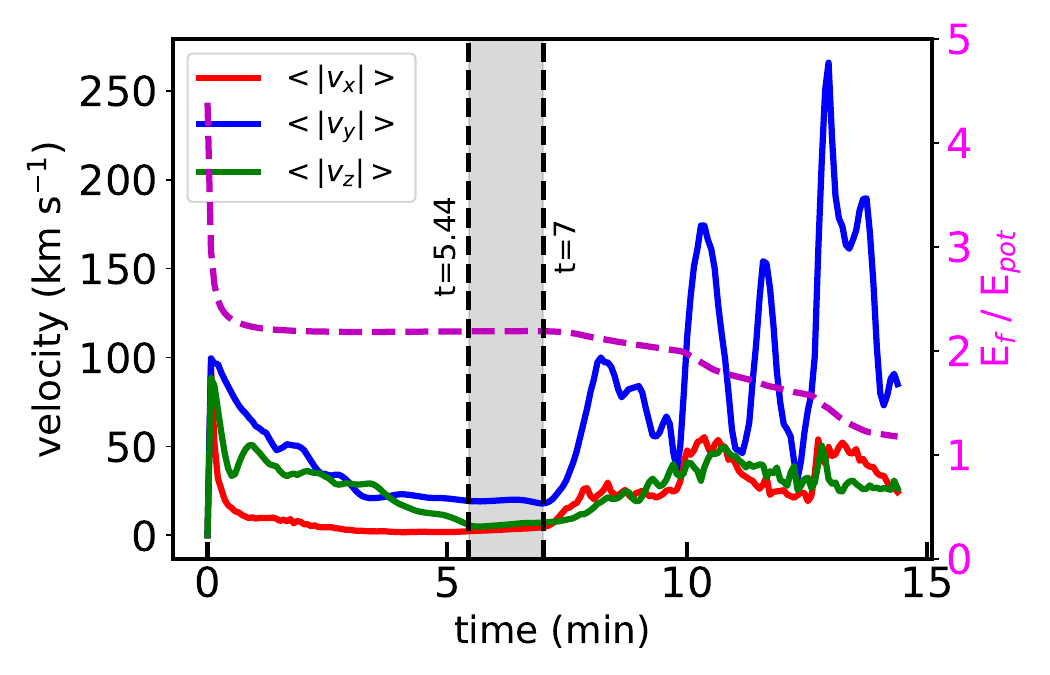}
   \caption{}
   \label{fig:semi_eq}
\end{subfigure}
\begin{subfigure}{0.88\columnwidth}
  \includegraphics[width=1\textwidth]{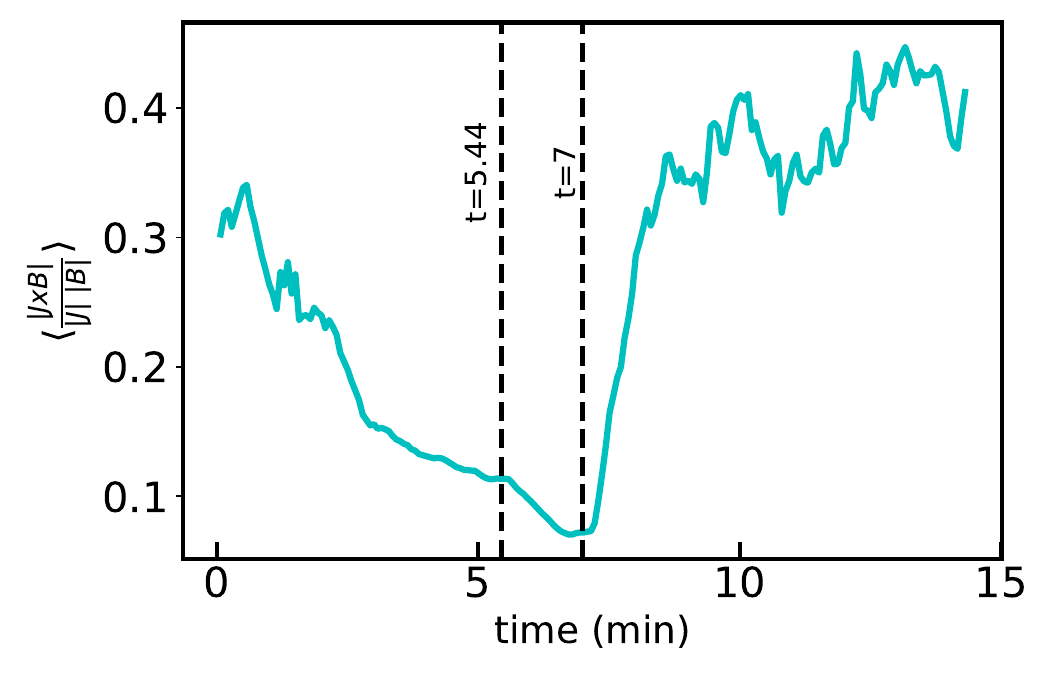}
   \caption{}
   \label{fig:ffness}
\end{subfigure}
\caption{(a) Temporal evolution of spatial averaged absolute velocity components $\langle |v_x(t)|\rangle$, $\langle |v_y(t)|\rangle$, $\langle |v_z(t)|\rangle$ and ratio of free ($E_f$) to potential ($E_{pot}$) energy (dashed curve in magenta). The shaded region between $t= 5.44$ and 7 min (shown by two vertical dashed lines) represents the dynamical semi-equilibrium phase of the system, where $\langle |v_y(t)|\rangle\ \approx 19$ km s$^{-1}$, $\langle |v_x(t)|\rangle\ \approx 2.2$ km s$^{-1}$, $\langle |v_z(t)|\rangle\ \approx 5.4$ km s$^{-1}$, and $E_f/E_{pot} \approx 2.1$. (b) Evolution of the spatial average, $\alpha = \langle \frac{|{\bf J} \times {\bf B}|}{|{\bf J}|\ |{\bf B}|}\rangle$.}
\label{fig:semi_equilb}
\end{figure}

\begin{figure*}[hbt!]
    \centering
\begin{subfigure}{0.33\textwidth}
  \includegraphics[width=1\textwidth]{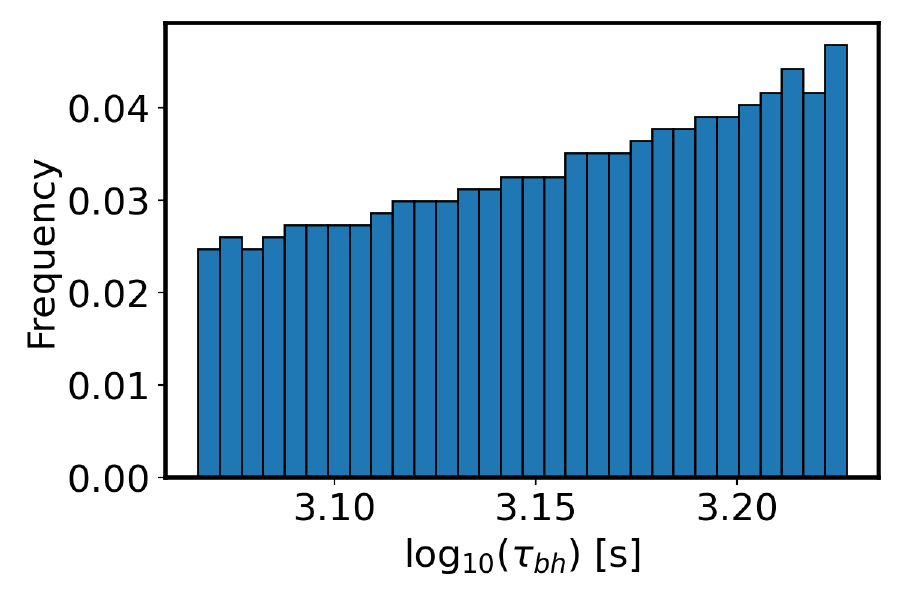}
   \caption{}
   \label{fig:bgr_ts}
\end{subfigure}
\begin{subfigure}{0.33\textwidth}
  \includegraphics[width=1\textwidth]{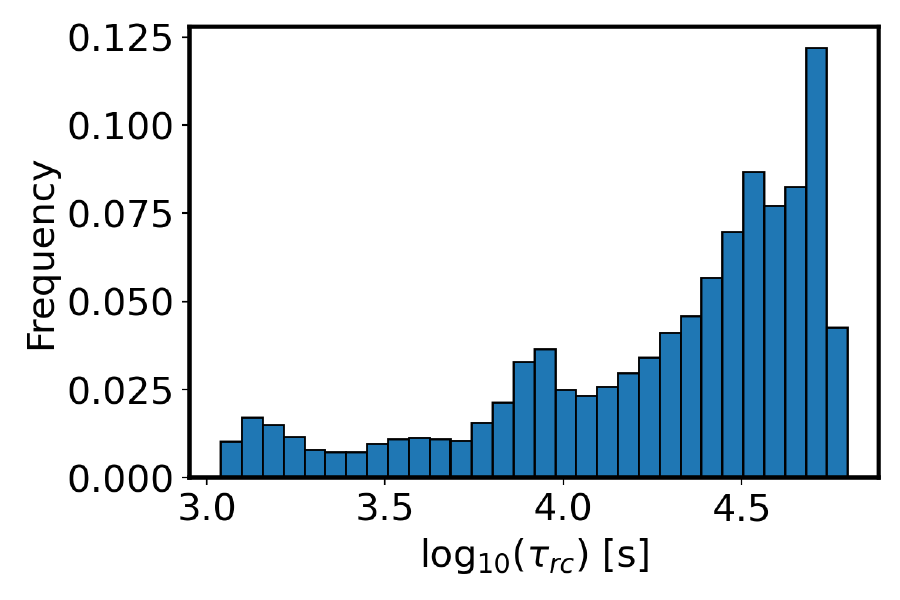}
   \caption{}
   \label{fig:rc_ts}
\end{subfigure}
\begin{subfigure}{0.33\textwidth}
  \includegraphics[width=1\textwidth]{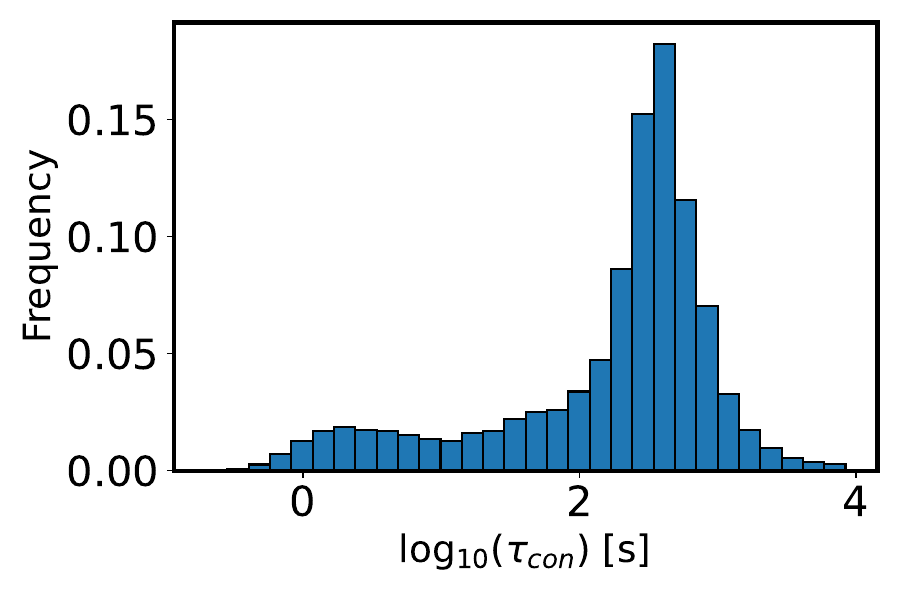}
   \caption{}
   \label{fig:con_ts}
\end{subfigure}
    \caption{Discrete probability densities of the characteristic time scales of (a) background heating ($\tau_{bh}$), (b) radiative cooling ($\tau_{rc}$), and (c) thermal conduction ($\tau_{con}$) at $t=5.44$ min.}
    \label{fig:cool_timescale}
\end{figure*}

\begin{figure*}
    \centering
    \includegraphics[width=0.85\textwidth]{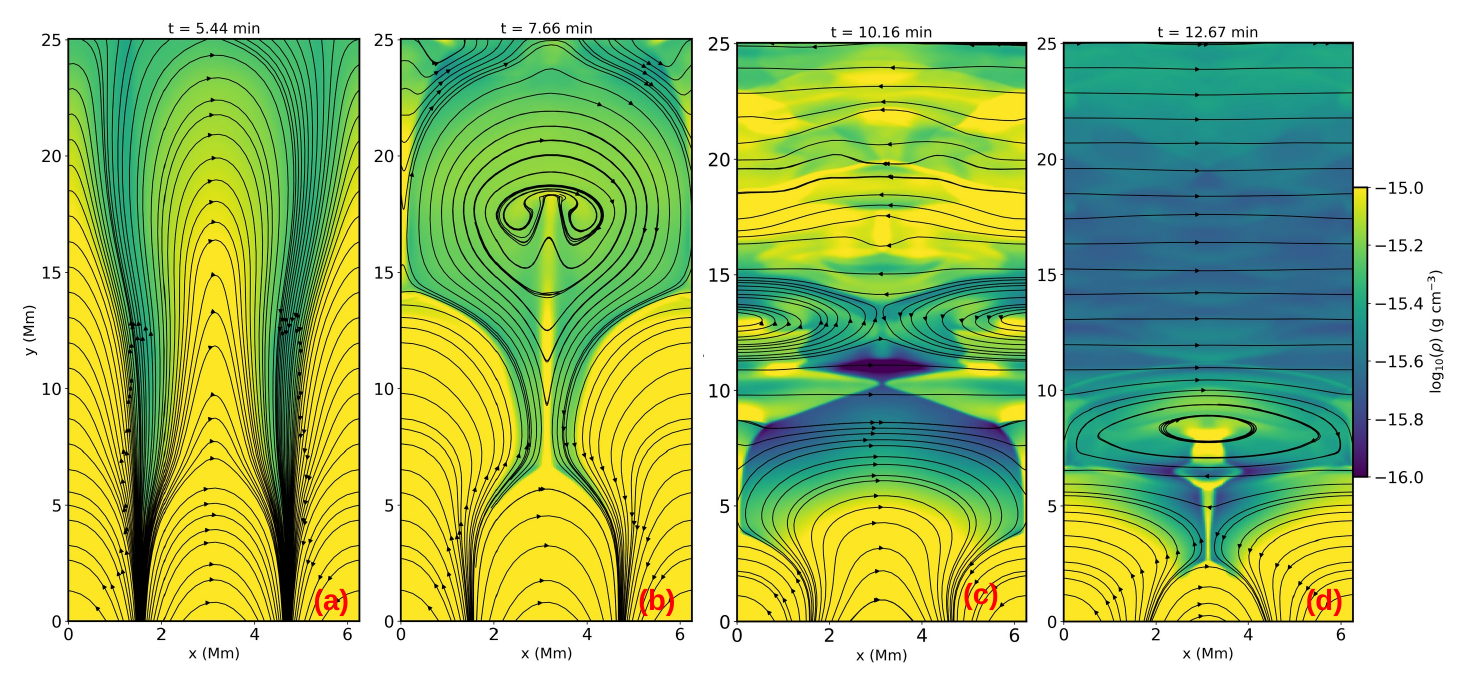}
    \includegraphics[width=0.85\textwidth]{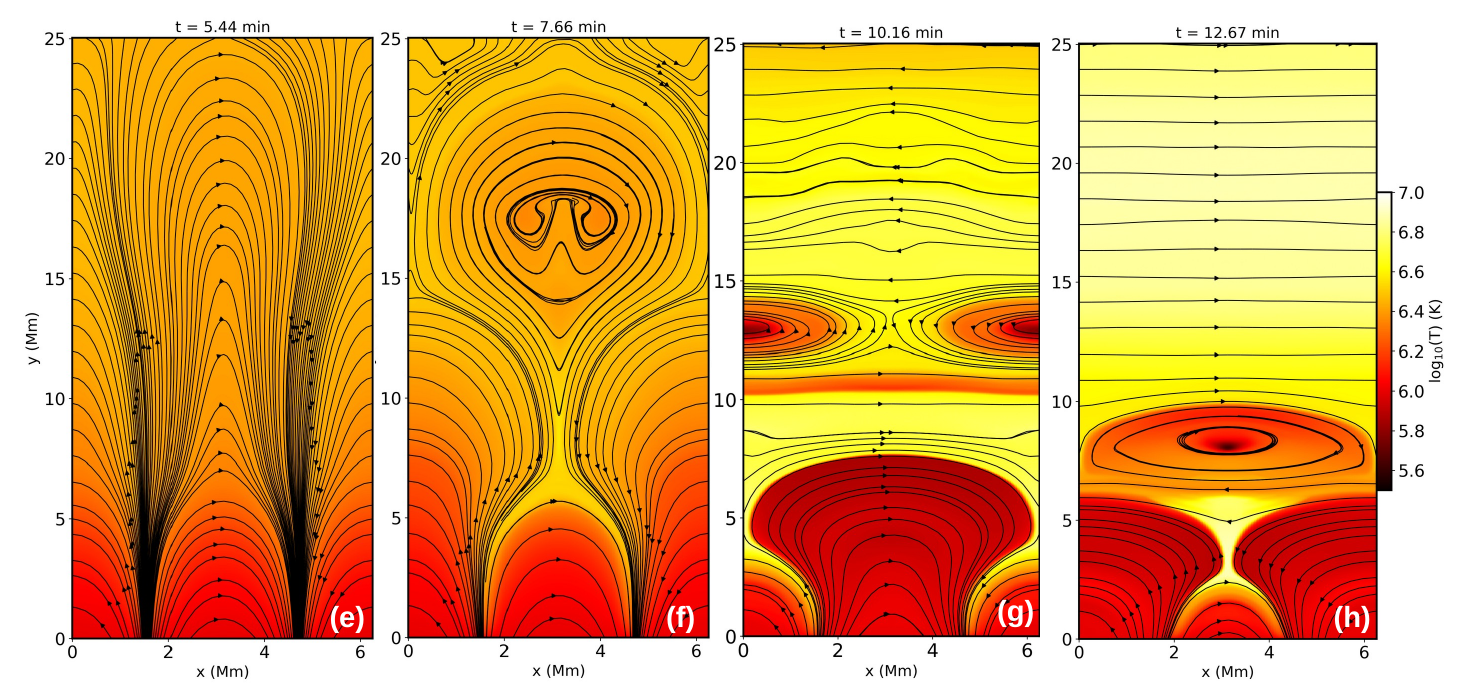}
    \includegraphics[width=0.85\textwidth]{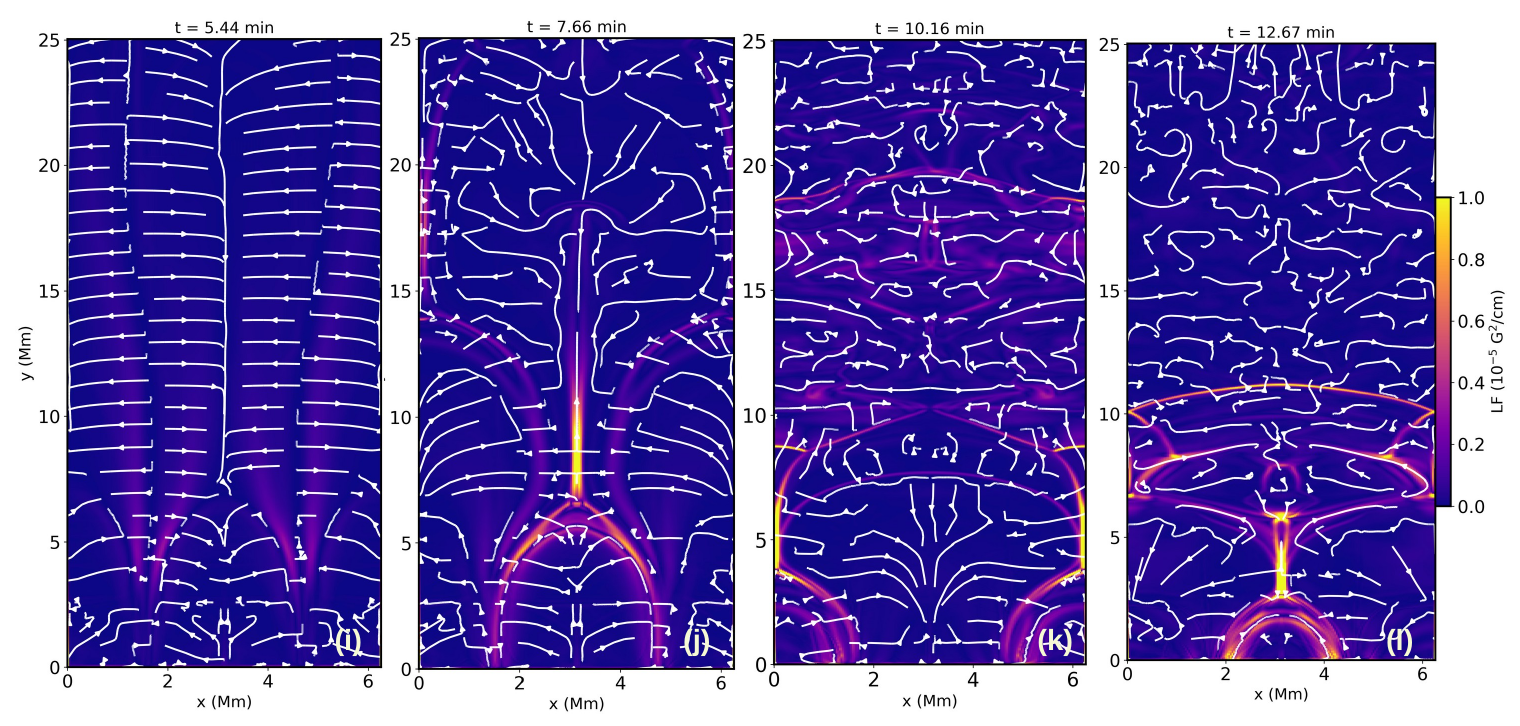}
    \caption{\textit{Top row:} Spatial distribution of plasma density for (a) semi-equilibrium stage. (b)-(d) shows the different MFR eruption stages, where the over-plotted solid curves represent the magnetic field lines projected in the $x-y$ plane. \textit{Middle row:} Same as the top panel for temperature. \textit{Bottom row:} Distribution of Lorentz force (LF) over the entire simulation domain, where the over-plotted arrows represent the LF direction in the $x-y$ plane. The saturation level of the colour bar is taken up to $10^{-5}$ G$^2$ cm$^{-1}$ for a better visualization of the enhanced LF regions. Animations of the figures are available online (height and width aspect ratio of the figures are not in scale).}
    \label{fig:den_early}
\end{figure*}

\begin{figure*}[hbt!]
\centering
\begin{subfigure}{\columnwidth}
  \includegraphics[width=1\textwidth]{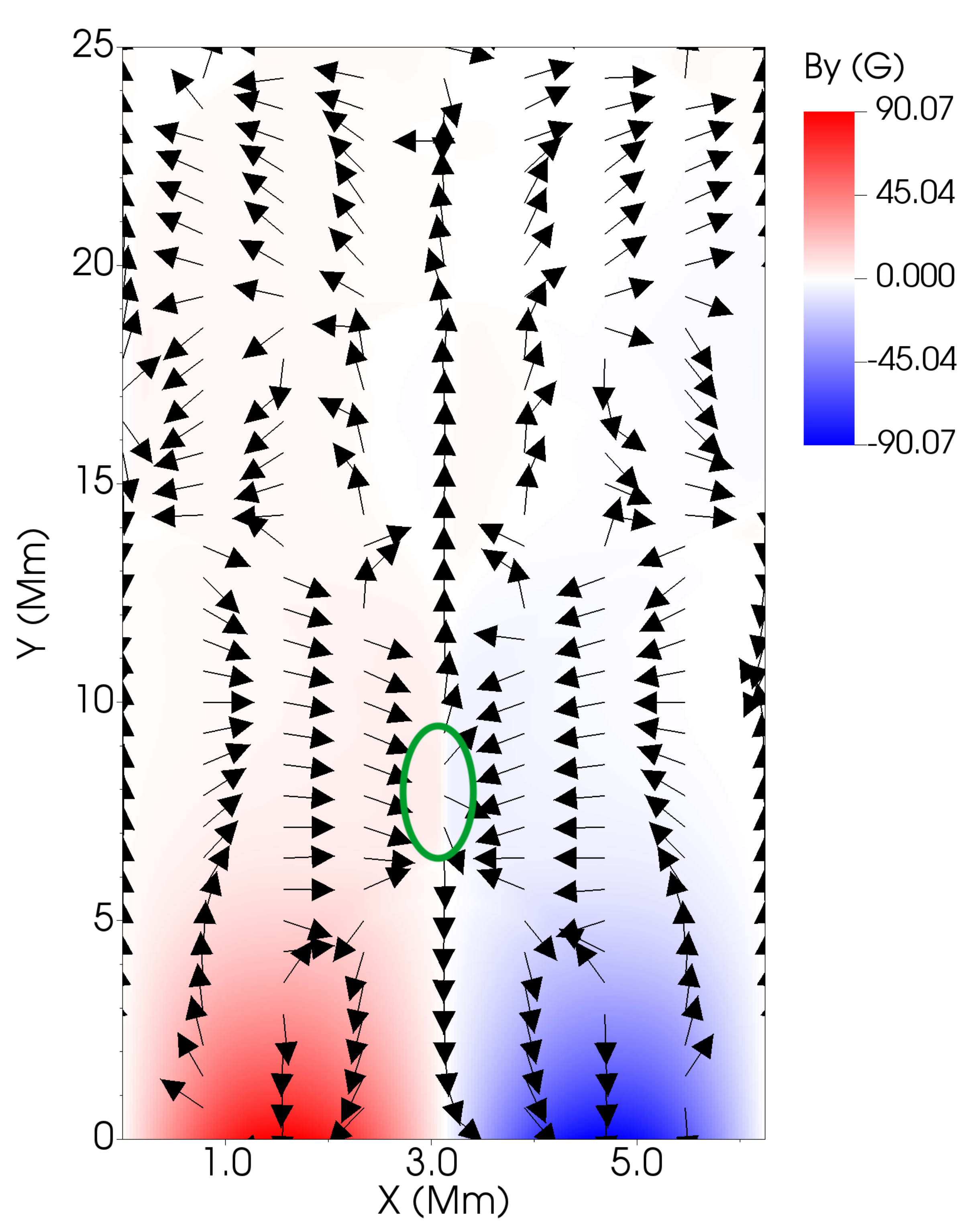}
   \caption{$t=7.66$ min}
   \label{fig:null1}
\end{subfigure}
\begin{subfigure}{\columnwidth}
  \includegraphics[width=1\textwidth]{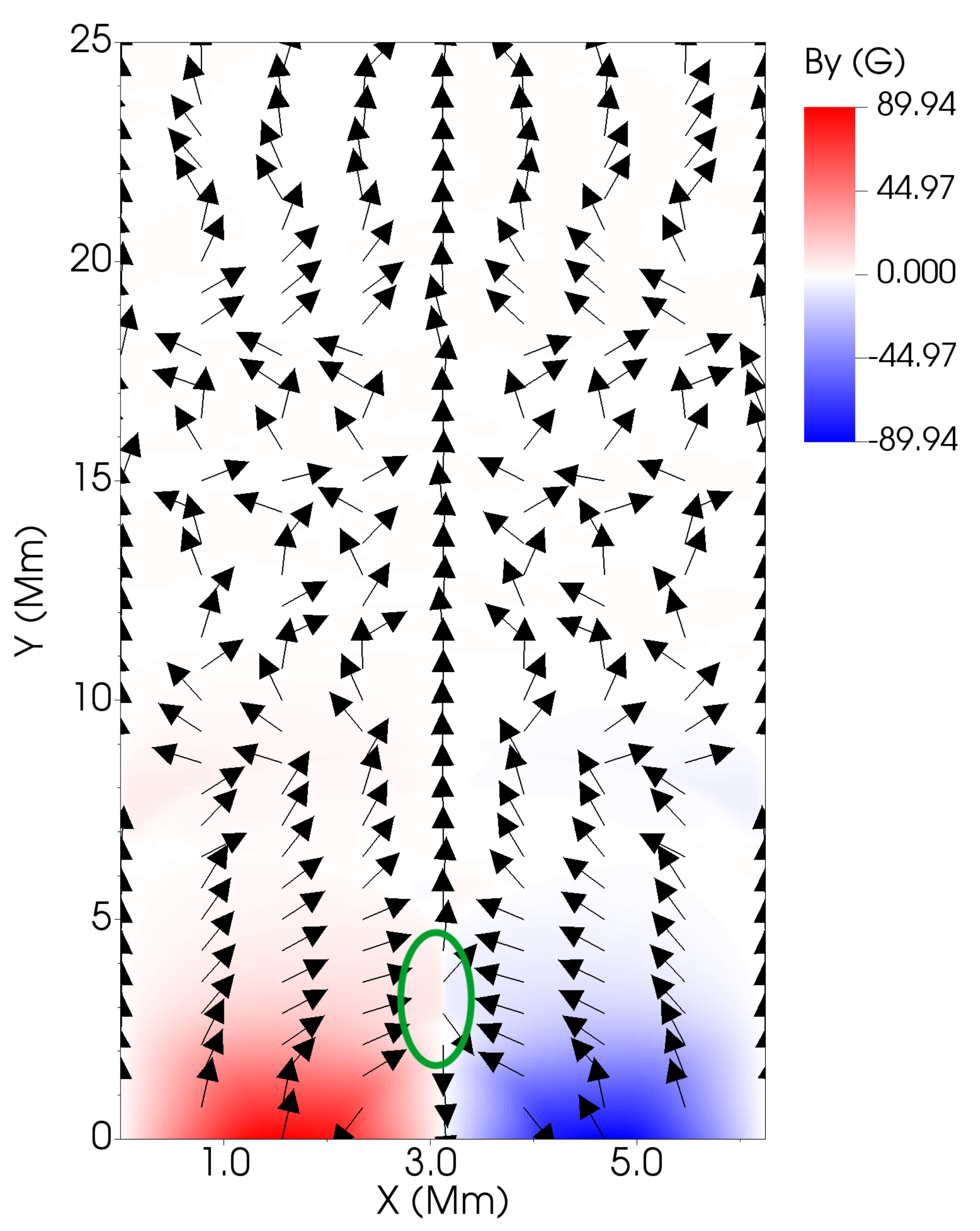}
   \caption{$t=12.67$ min}
   \label{fig:null2}
\end{subfigure}

\begin{subfigure}{\columnwidth}
  \includegraphics[width=1\textwidth]{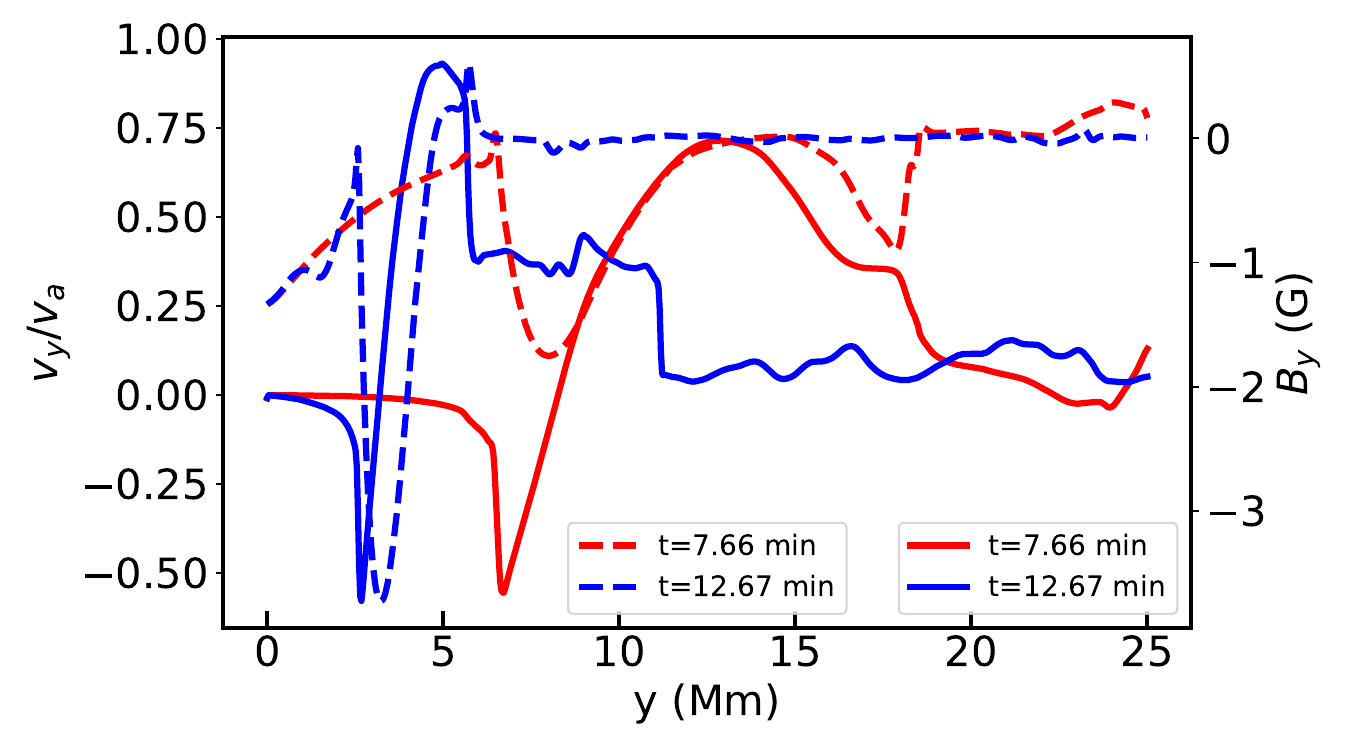}
   \caption{$x=3.14$ Mm}
   \label{fig:rec_velo_y}
\end{subfigure}
\begin{subfigure}{\columnwidth}
  \includegraphics[width=0.9\textwidth]{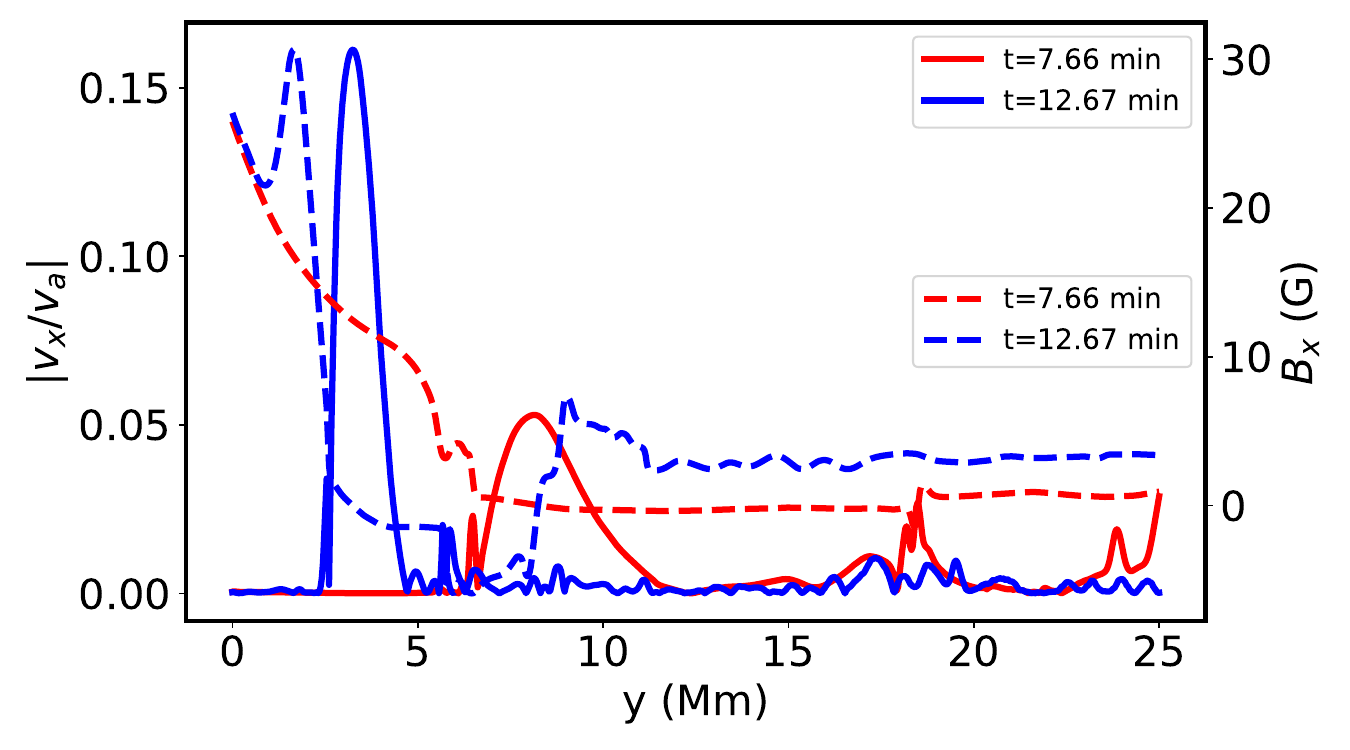}
   \caption{$x=3.14$ Mm}
   \label{fig:rec_velo_x}
   \end{subfigure}
\caption{Variation of $B_y$ field in the $x-y$ plane is shown at (a) $t=7.66$ and (b) 12.67 min, where the over-plotted black arrows represent the velocity field $\lbrace v_x, v_y \rbrace$ in the $x-y$ plane. The region inside the green ellipses (in Figures (a) and (b)) contain the reconnection point locations, where we see the plasma inflow and outflow due to reconnection (height and width aspect ratio in the figures are not in scale). (c) $v_y$ and $B_y$ distributions along $y$-direction at $x=3.14$ Mm (which covers the reconnection point location). (d) Absolute velocity of $v_x$ and $B_x$ distributions along $y$-direction at $x=3.14$ Mm for the eruption phases. The velocities are scaled in units of instantaneous Alfv{\'e}n speed ($v_a$). The solid and dashed curves in Figures (c) and (d) represent the velocities and magnetic field strengths respectively.} 
\label{fig:null}
\end{figure*}

\begin{figure}[hbt!]
    \centering
    \includegraphics[width=0.9\columnwidth]{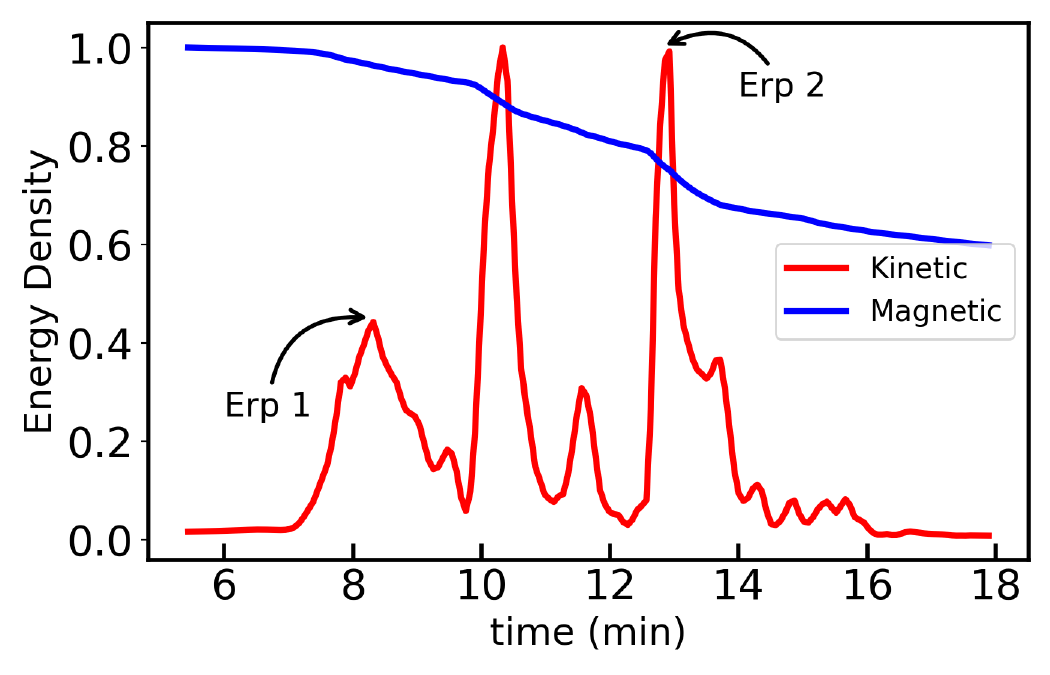}
    \caption{Temporal evolution of (normalized) mean kinetic and magnetic energy densities during the eruption phases, where the black arrows represent the time when first (Erp 1) and second (Erp 2) eruptions from the central region have the peak KE densities. The other peaks between the two eruptions (Erp 1 and 2) show the signature due to eruptions from the side boundaries.}
    \label{fig:eruption}
\end{figure}

\begin{figure}[hbt!]
\centering
\begin{subfigure}{\columnwidth}
  \includegraphics[width=1\textwidth]{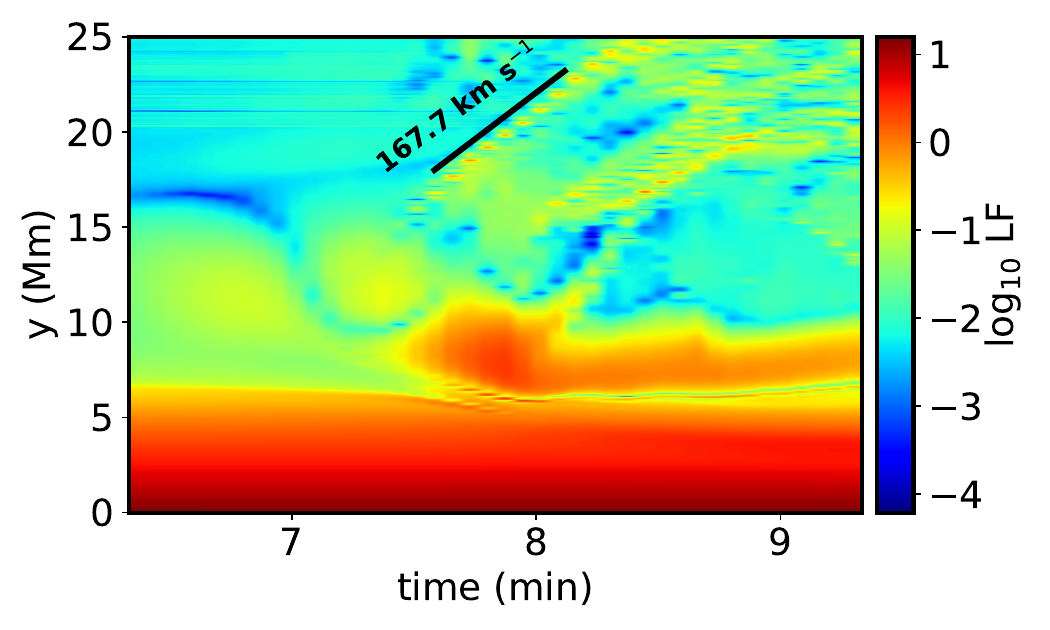}
   \caption{}
   \label{fig:lf_ts1}
\end{subfigure}
\begin{subfigure}{\columnwidth}
  \includegraphics[width=1\textwidth]{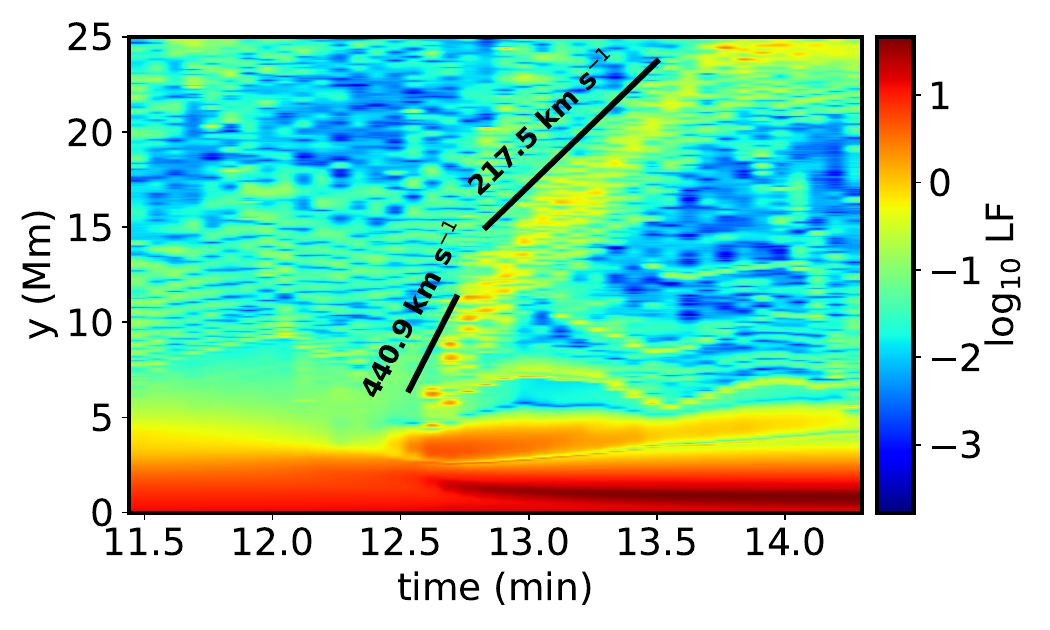}
   \caption{}
   \label{fig:lf_ts2}
\end{subfigure}
\caption{Time-distance map of LF (in arbitrary unit) along the vertical cut at $x=3.14$ Mm. (a) and (b) represent the time domain around the first and second flux rope eruption phases, respectively. The slope(s) at the selected part(s) in the maps are highlighted by the black lines to estimate the eruption speed of the flux ropes.} 
\label{fig:lf_ts}
\end{figure}

\begin{figure*}[hbt!]
\centering
\begin{subfigure}{0.33\textwidth}
  \includegraphics[width=1\textwidth]{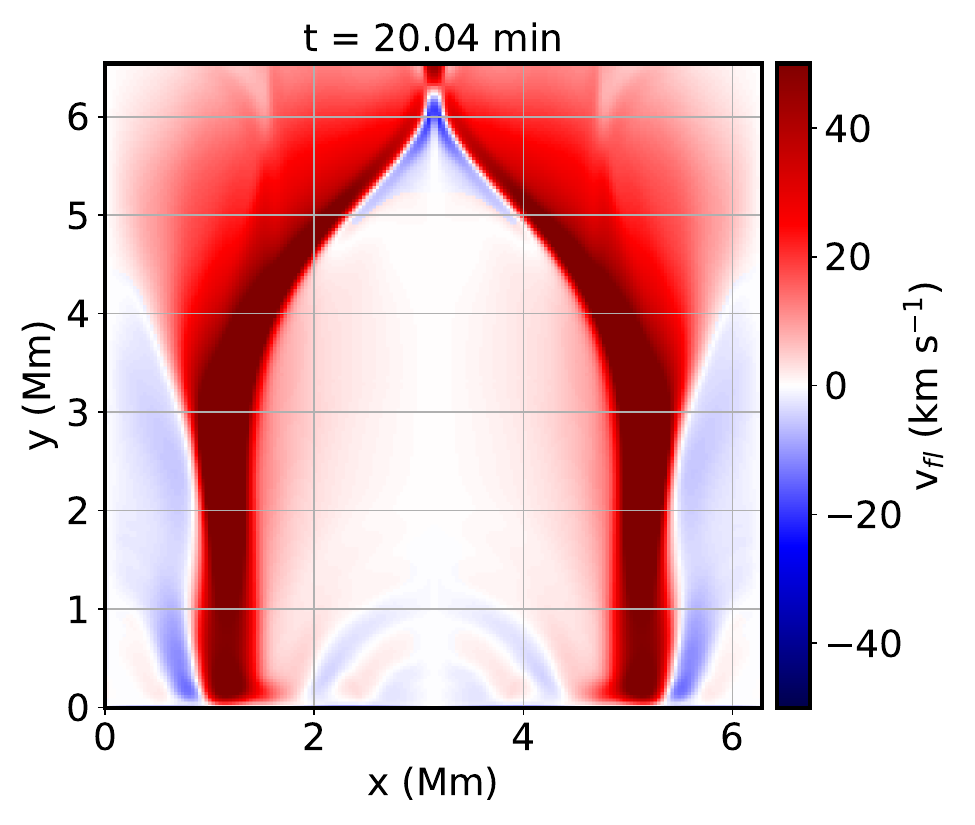}
   \caption{}
   \label{fig:vfl280}
\end{subfigure}
\begin{subfigure}{0.33\textwidth}
  \includegraphics[width=1\textwidth]{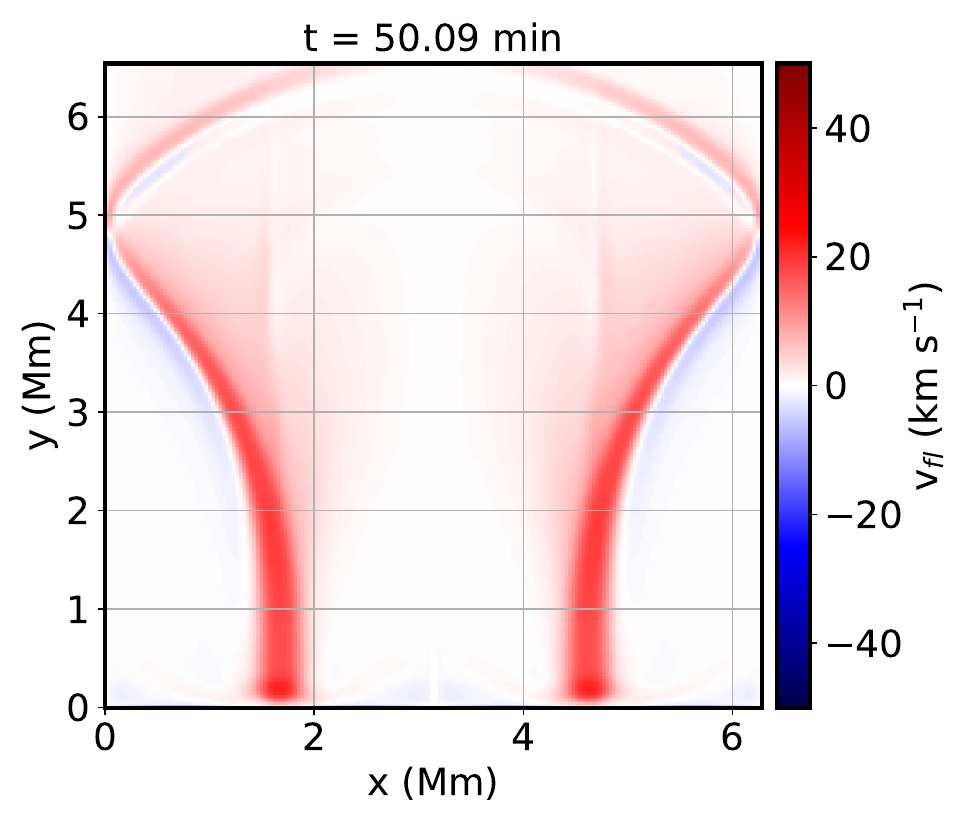}
   \caption{}
   \label{fig:vfl700}
\end{subfigure}
\begin{subfigure}{0.33\textwidth}
  \includegraphics[width=1\textwidth]{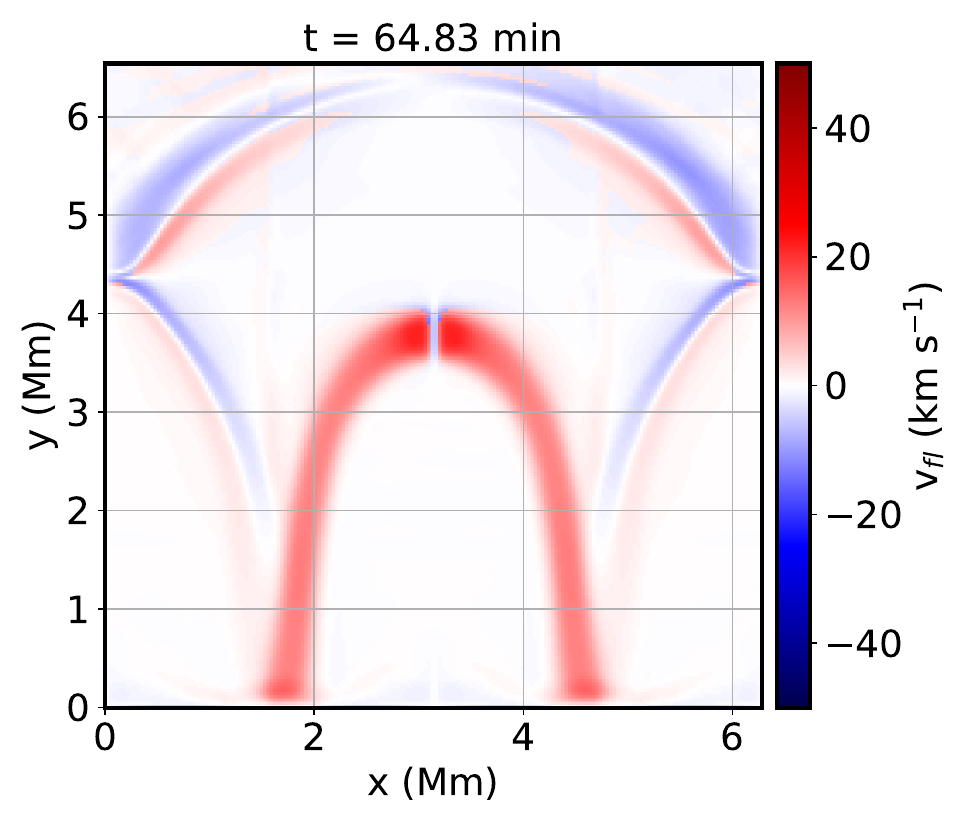}
   \caption{}
   \label{fig:vfl906}
\end{subfigure}
\caption{Velocity map along the projected field lines on the vertical $x-y$ plane, $v_{fl}$, for three different times.} 
\label{fig:vfl}
\end{figure*}

\begin{figure}[hbt!]
    \centering
    \includegraphics[width=0.4\textwidth]{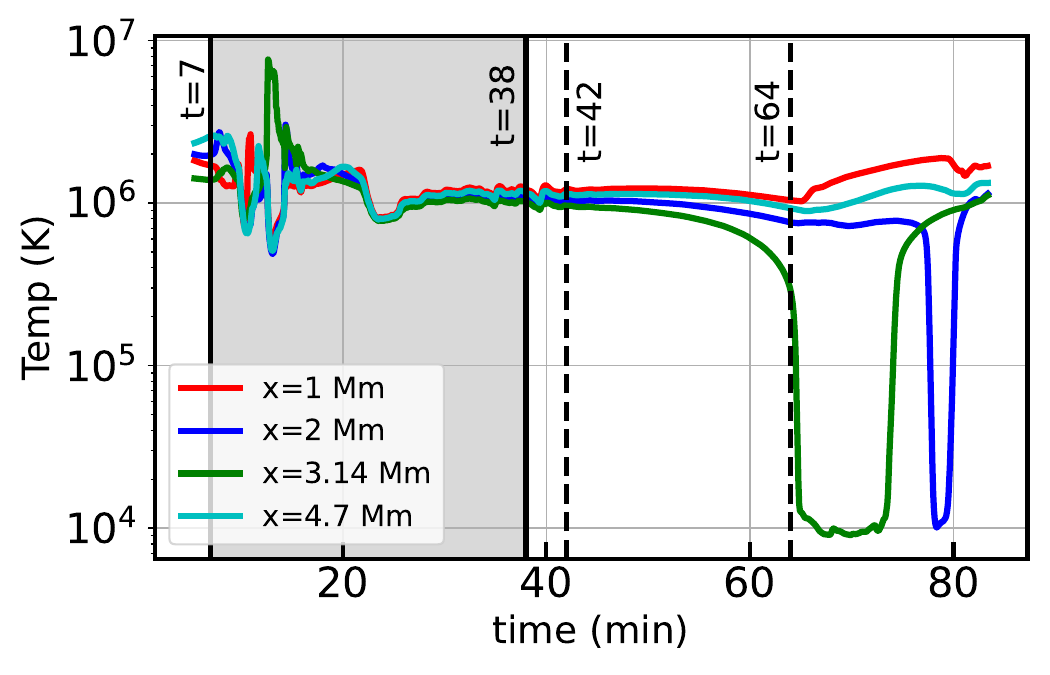}
    \includegraphics[width=0.4\textwidth]{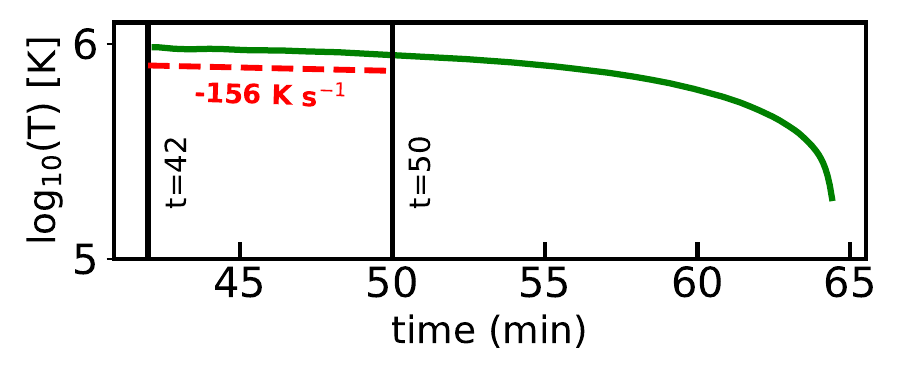}
    \caption{\textit{Top panel:} The temperature evolution at four different horizontal locations at $x=1, 2, 3.14$, and 4.7 Mm all at the same height $y=4$ Mm. The shaded region between $t=7$ to 38 min represent the eruption phase, where several MFRs leave the domain. The vertical dashed line at $t=42$ min marks the time when the upper coronal part again settles down to around 1 MK. The dashed line at $t=64$ min represents the onset of a thermal runaway process, which happens first in the central $x=3.14$ Mm location. \textit{Bottom panel:} Temperature evolution in the central $(x, y)=(3.14, 4)$ Mm location. The temperature drops between $t=42$ and 50 min with a decay rate of 156 K s$^{-1}$ as represented by the red dashed line.}
    \label{fig:flare-duration}
\end{figure}

\begin{figure}[hbt!]
    \centering
    \includegraphics[width=0.9\columnwidth]{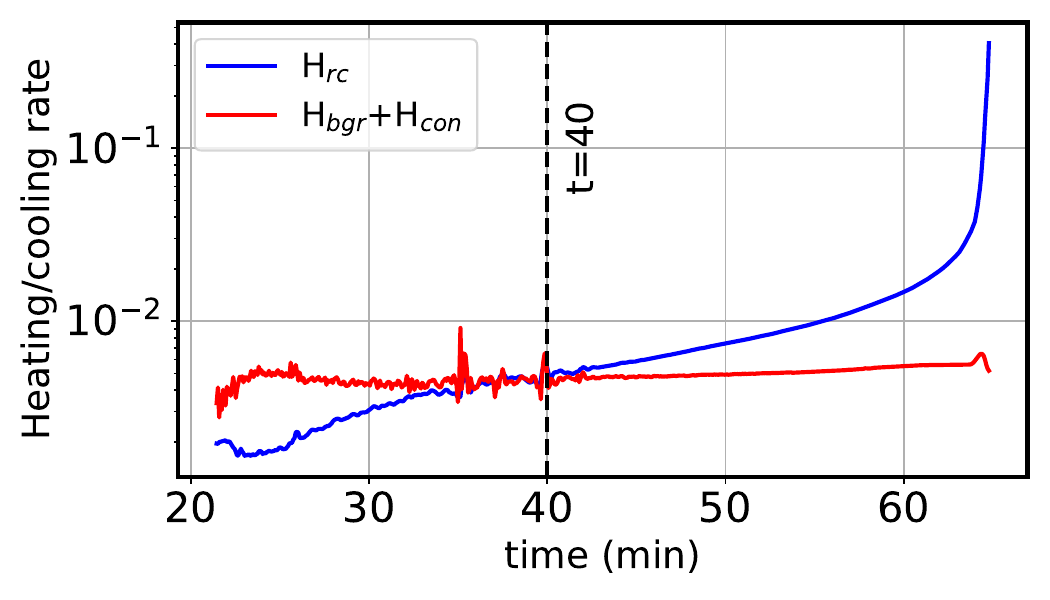}
    \caption{The development of thermal imbalance between heating and cooling at the loop top location $(x, y) = (3.14, 4)$ Mm in the late phase of our simulation. The red curve denotes the heating rate due to thermal conduction ($H_{con}$) and background heat ($H_{bgr}$), and the blue curve denotes the cooling rate due to radiation ($H_{rc}$). The units for all the quantities are in erg cm$^{-3}$ s$^{-1}$. The vertical dashed line marks the time after which the radiative cooling starts to dominate over the heating due to conduction and background heat.}
    \label{fig:heating-cooling-misbalance}
\end{figure}

\section{Results and analysis}\label{sec:results}

\subsection{Achieving semi-equilibrium phase of the system}\label{sce:semi-equilibrium}

Due to the initial non-force-free nature of the magnetic field configuration, and the corresponding mechanical imbalance, the system is forced to evolve away from this state. We follow the evolution of the spatial average of the absolute velocity components,
\begin{align}
    \langle v_{x,y,z} \rangle = \frac{1}{V} \int_V |v_{x,y,z}| \ {\rm d}V
\end{align}
over the entire simulation domain, $V$ as shown in Figure \ref{fig:semi_eq}. The average velocity components of the system take a jump due to the large Lorentz force from the initial state ($t=0$). The maximum velocity jump of the $v_y$ component, which is $\approx 100$ km s$^{-1}$, whereas the other two components are smaller. Then the system relaxes achieving a minimum velocity $ \langle |v_y| \rangle \approx$ 19 km s$^{-1}$ at $t=5.44$ min which remains almost constant till $t=7$ min. The other components, $\langle |v_x| \rangle$, and $\langle |v_y| \rangle$ also remain constant to $\approx 2.2$ and 5.4 km s$^{-1}$ respectively within this time domain. This is marked by the shaded region in Figure \ref{fig:semi_eq}. To estimate the energetics of this rapid initial relaxation phase of the evolution, we calculate the ratio of the free and potential field energies at each time steps between $t=0$ to 15 min. The potential field energy ($E_{pot}$) is calculated as 
\begin{align}
    E_{pot} = \int_V \frac{B_{pot}^2}{2} {\rm d}V,
\end{align}
where $\vec{B}_{pot}$ is the potential field extrapolated using the instantaneous $B_y$ component at the base of the simulation domain ($y=0$). The free energy $E_f$ is the excess magnetic energy of the system from its corresponding potential field energy, given by
\begin{align}
    E_f = \int_V \frac{B^2}{2} {\rm d}V - E_{pot} \,. 
\end{align}

We notice from Figure \ref{fig:semi_eq} that the value of $E_f/E_{pot}$ decreases sharply from 4.38 at the initial state to about 2.1 within $\approx 1$ min, which remains constant till 7 min. The reason why we see an evolution of the ratio $E_f/E_{pot}$ towards the value two can be argued from changes in the magnetic energy observed in active regions during major flares \citep{2021A&A...653A..69G, 2023ApJ...942...27L}. To estimate the evolution of the Lorentz force, we calculate the spatial average of the parameter,

\begin{align}
    \alpha = \frac{1}{V} \int_V \frac{|{\bf J} \times {\bf B}|}{|{\bf J}|\ |{\bf B}|} \ {\rm d}V,
\end{align}
 within the full simulation domain at each time step as shown in Figure \ref{fig:ffness}. This parameter represents the average relative angle formed by the current and the magnetic field which is useful for determining the (non-)force-free nature of the magnetic field. The initial value of the average angle between ${\bf J}$ and $\bf{B}$ is $\arcsin(\alpha) \approx 18^\circ$ which reduces to $\approx 5^\circ$ at $t\approx 5.44$ min, and has a minimum value of $\approx 4^\circ$ at $t=7$ min. This implies that the system achieves a nearly force-free state within minutes, and that the period between 5 and 7 minutes corresponds to a nearly semi-equilibrium phase. We will explain further on how the changes after $t=7$ min are accompanied by large flux rope expulsions and flare-loop configurations, which again cause large velocity and Lorentz force variations. We first characterize the thermodynamical timescales for the nearly-force-free state between $t=5.44$ and 7 min.   
 
We estimate the characteristic time scales of background heating ($\tau_{bh}$), radiative cooling ($\tau_{rc}$), and field-aligned thermal conduction ($\tau_{con}$) at the starting time of the dynamical semi-equilibrium state at $t=5.44$ min, which are calculated by 
\begin{align}
    \tau_{bh} &= \frac{\epsilon}{H_{bgr}} \,,\\
    \tau_{rc} &= \frac{\epsilon}{\rho^2 \Lambda(T)} \,,\\
    \tau_{con} &= \frac{\epsilon}{|\nabla \cdot (\kappa_{||}{\bf b} ({\bf b} \cdot \nabla) T)|}\,,
\end{align}
respectively, where $\epsilon = p/(\gamma-1)$ is the internal energy density per unit volume. We calculate these characteristic time scales at each pixel of the simulation domain, and obtain probability densities for them, as shown in Figure \ref{fig:cool_timescale}. We notice that $\tau_{bh}$ and $\tau_{rc}$ are more than 1000 s for the entire simulation domain (see Figures \ref{fig:bgr_ts} and \ref{fig:rc_ts}). For estimating the (magnetic field aligned) thermal conduction time scales $\tau_{con}$, we use a convolution filter using a uniform kernel of size $5 \times 5$ to smooth out numerical discretization bias that arises in specific regions. We notice that $\approx 76\%$ of the domain corresponds to $\tau_{con} > 100$ s, and only 1.05\% of the domain has less than 1 s. However, the dynamical semi-equilibrium phase of the system lasts for $\approx 93$ s (the shaded region in Figure \ref{fig:semi_eq}). Taking all these aspects into consideration, where cooling and heating times are long and of similar magnitude, while thermal conduction quickly equilibrates temperatures along loops, we argue that the semi-equilibrium phase between 5.44 and 7 min is a relevant starting point to study the later evolution of the system.

\subsection{Development and eruption of flux ropes} \label{sec:fluxrope}

The plasma density, temperature, Lorentz force, and magnetic field line distribution of the semi-equilibrium state (at $t=5.44$ min) are shown in the first column of Figure \ref{fig:den_early}. The magnetic field lines resemble a coronal loop structure embedded in a bipolar magnetic field configuration. The evolution during the semi-equilibrium phase is slow in comparison to the violent initial phases and the thermal processes are in no position to substantially modify the plasma during that phase; yet it is not a pure equilibrium; rather, there is a small imbalance in the forces which leads to the events described in the following. 

The LF tries to squeeze the magnetic arcades along the horizontal direction near the central region at $x=3.14$ Mm (and also towards the periodic horizontal boundaries). This is clearly noticeable in Figure \ref{fig:den_early}(i) for the upper $y \gtrapprox 7$ Mm region. The squeezing of opposite polarity field lines creates a localized region with small (of order one Gauss) $B_x$ and $B_y$ components in the presence of the guide field, $B_z$. This location leads to the magnetic reconnection without null point due to the presence of (uniform) resistivity \citep{1995JGR...10023443P}, which can be noticed in Figures \ref{fig:den_early}(b) and \ref{fig:den_early}(f). At around $\approx 7.66$ min, we see the formation of a vertical current sheet (CS) in the central region between $y\approx 5-10$ Mm, where the LF is enhanced, as shown in Figure \ref{fig:den_early}(j). This CS region becomes hot up to $\approx 4$ MK as shown in Figure \ref{fig:den_early}(f). Due to reconnection in the CS plane, an erupting magnetic flux rope (MFR) is formed in the central region (see Figure \ref{fig:den_early}(b) or \ref{fig:den_early}(f)). Due to the periodic condition at the horizontal boundaries, erupting MFRs are also formed later at $\approx 10$ min at the side boundaries. This is shown in Figs. \ref{fig:den_early}(c) and \ref{fig:den_early}(g). Further spontaneous reconnections lead to the formation of a second erupting flux rope in the central region at a lower height between $y \approx 0-5$ Mm at around $\approx 12.67$ min (see Figs \ref{fig:den_early}(d) and \ref{fig:den_early}(h)), again accompanied by the formation of a CS underneath it. The temperature of this CS rises to $\approx 7$ MK due to resistive heating. Cusp like structures are formed beneath the vertical CSs as shown in Figures \ref{fig:den_early}(b), (d), (f), and (h). This is similar to the classic CSHKP solar flare model \citep{C, S, H, K}, where hot CSs are formed below erupting MFRs, and cusp like magnetic structures form underneath CSs.

We show the flow patterns and the $B_y$ distributions for the instants with clear central MFR eruptions in Fig.~\ref{fig:null1}-\ref{fig:null2}. The reconnection location in the bipolar magnetic structure is highlighted by the green ellipses in Figures \ref{fig:null1}, and \ref{fig:null2}, where the reconnection-driven inflows and outflows are evident. To appreciate the reconnection location, we quantify the distribution of $v_y$ and $B_y$ shown in Figure \ref{fig:rec_velo_y}, and $v_x$ and $B_x$ in Fig. \ref{fig:rec_velo_x} by taking a vertical cut at $x=3.14$ Mm at $t=7.66$ and $12.67$ min respectively. The regions where the outflow velocities flip directions are spatially co-located with the small (of order one Gauss) magnetic field strength locations of $B_x$ and $B_y$. The reconnection points are located in the central region ($x=3.14$ Mm) at $y\approx 7.5$ Mm for the first, and $y\approx 3$ Mm for the second (central) eruptions. The maximum outflow velocities remain sub-Alfv{\'e}nic, where the Alfv{\'e}n velocity ($v_a$) is calculated based on the instantaneous local plasma density and magnetic field strength at a fixed location in the vicinity of the inflow side near the reconnection points. To estimate the reconnection rate at those locations we consider the ratio of the magnitude of inflow velocity with respect to the instantaneous Alfv{\'e}n velocity, $\mathcal{M}=|v_x/v_a|$. The variation of $|v_x/v_a|$ in the central region is shown in Figure \ref{fig:rec_velo_x}, where we see the reconnection rate is $\approx 0.05$ for the first eruption, and $\approx 0.15$ for the second eruption near the reconnection points. It implies that the second eruption is more violent than the first one. 

This MFR eruption phase, where successive flux ropes are shed from the domain and escape through the top boundary, which is open to outflows, is energetically quantified in Fig.~\ref{fig:eruption}. Temporal distributions of the mean kinetic (KE) and magnetic (ME) energy densities, which are calculated by the spatial average of the entire simulation domain,
\begin{align}\label{eq:ke}
    \text{KE} &= \frac{1}{V} \int_V \frac{\rho v^2}{2} {\rm d}V,\\ \label{me}
    \text{ME} &= \frac{1}{V} \int_V \frac{B^2}{2} {\rm d}V,
\end{align}
are shown in Figure \ref{fig:eruption}. The KE exhibits strong peaks two of which occur at around $\approx 8$ and 13 min and are associated with the first and second flux rope eruptions (marked as `Erp 1' and `Erp 2' respectively) from the central region. The other peaks (one strong and one relatively weaker) in between the two central eruptions are due to the secondary eruptions occurring from the side boundaries (the first one is shown in Figure \ref{fig:den_early}(c); the second one is not shown in the figure, but can be found in the associated animation available online). During the eruptions, the magnetic energy density decreases as free magnetic energy is converted into kinetic (and thermal) energy, and thus we see the corresponding drop of the magnetic energy density. The drops are sharper during the peak eruption times when the KE density has peaks as shown in Figure \ref{fig:eruption}. We see clear signatures in Figures \ref{fig:den_early}(j), and \ref{fig:den_early}(l) that the enhanced LF regions are co-located with the MFR locations of Figures \ref{fig:den_early}(b), and \ref{fig:den_early}(d) respectively. 

To explore the dynamics of the eruptions themselves, we produce time-distance (TD) maps of LF along a vertical cut at $x=3.14$ Mm, as shown in Figure \ref{fig:lf_ts}. Note that the choice of LF instead of density (or temperature) is preferred in this case as the LF maps show a clear signature of the MFRs location. For analysing the first eruption, we select the temporal domain of the TD map between $\approx 6.3$ to 9.33 min, as shown in Figure \ref{fig:lf_ts1}, where we track the enhanced LF region in the appropriate spatial and temporal domain corresponding to the flux rope eruption, and estimate the eruption speed by calculating the slope (shown by the black solid line) of that region in the $y-t$ plane. Similarly, for the second eruption (from the central region), we select the temporal domain between $\approx 11.4$ to 14.33 min, as shown in Figure \ref{fig:lf_ts2}, and track the enhanced LF regions corresponding to the eruption to estimate the eruption speeds. The speed of the first eruption is estimated to be $\approx 167$ km s$^{-1}$. For the second eruption, the flux rope speed is $\approx 441$ km s$^{-1}$ at the early phase and then decelerates to $\approx 217$ km s$^{-1}$. This is in fair agreement with observational evidence in solar flares, where the average rising velocity of the flux rope is reported $\approx 224$ km s$^{-1}$ \citep{2018ApJ...853L..18Y}. The density is decreased in the ambient medium at around $y \approx 10-20$ Mm after the first eruption (see Figure \ref{fig:den_early}(d)). Also, the formation of the second flux rope occurs at a lower height with a higher LF. These two effects make the eruption of the second MFR faster than the first one. This is also reflected in Figure \ref{fig:eruption}, where we see that the mean kinetic energy peak for the second eruption (Erp 2) is more than the peak value of the first eruption (Erp 1).

\begin{figure*}[hbt!]
\centering
\begin{subfigure}{0.3\textwidth}
  \includegraphics[width=1\textwidth]{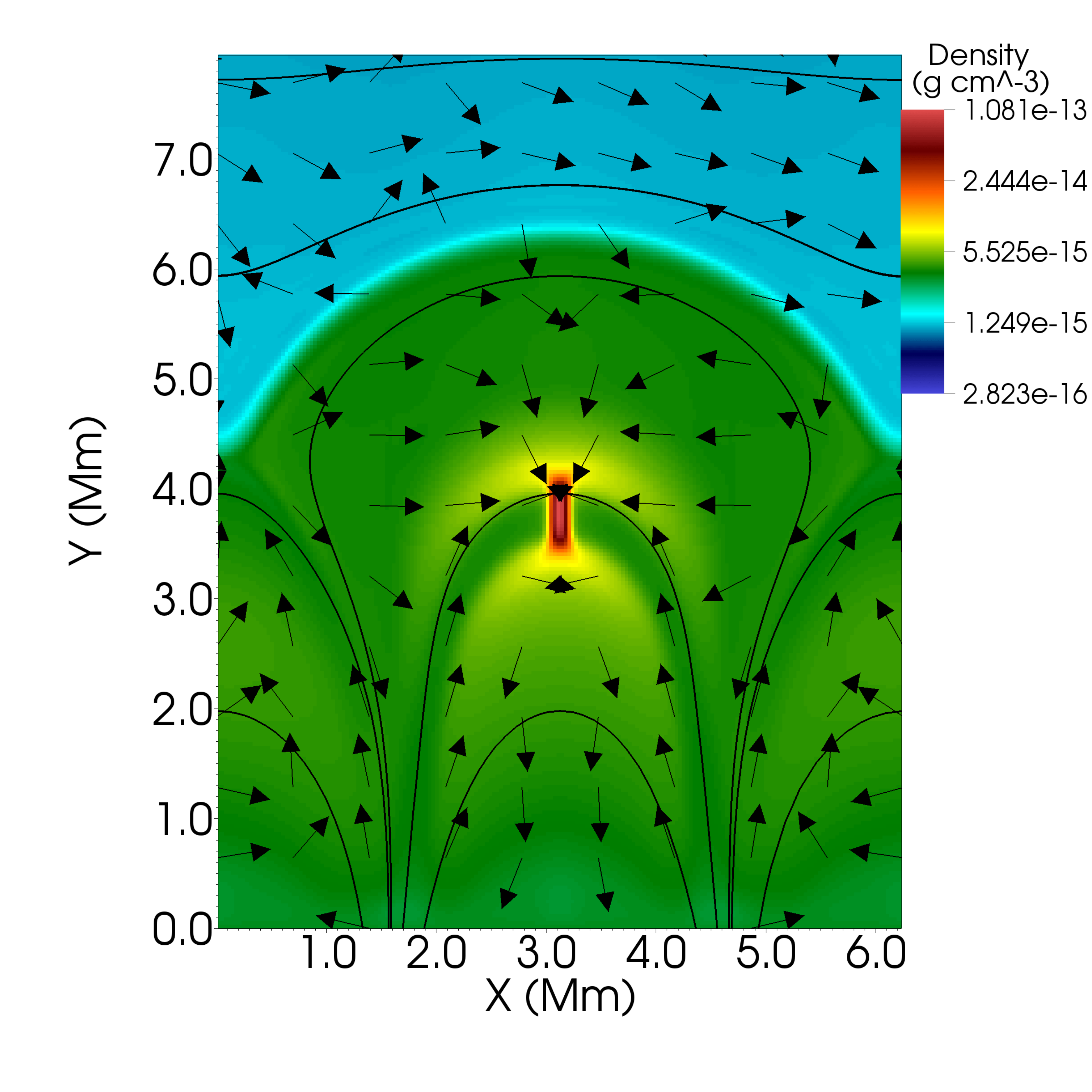}
   \caption{$t=64.83$ min}
   \label{fig:den_t0906}
\end{subfigure}
\begin{subfigure}{0.3\textwidth}
    \includegraphics[width=1\textwidth]{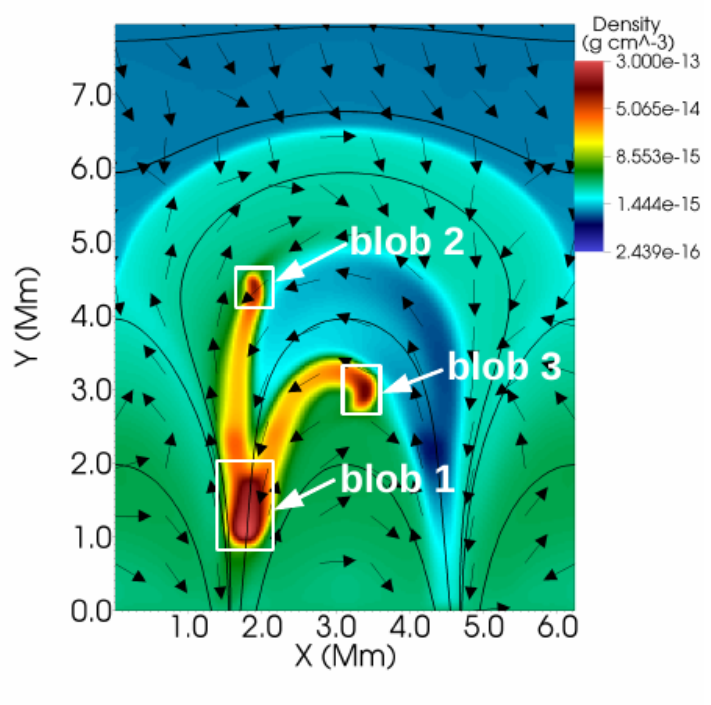}
    \caption{$t=80.10$ min}
    \label{fig:den_t1120}
\end{subfigure}
\begin{subfigure}{0.3\textwidth}
    \includegraphics[width=1\textwidth]{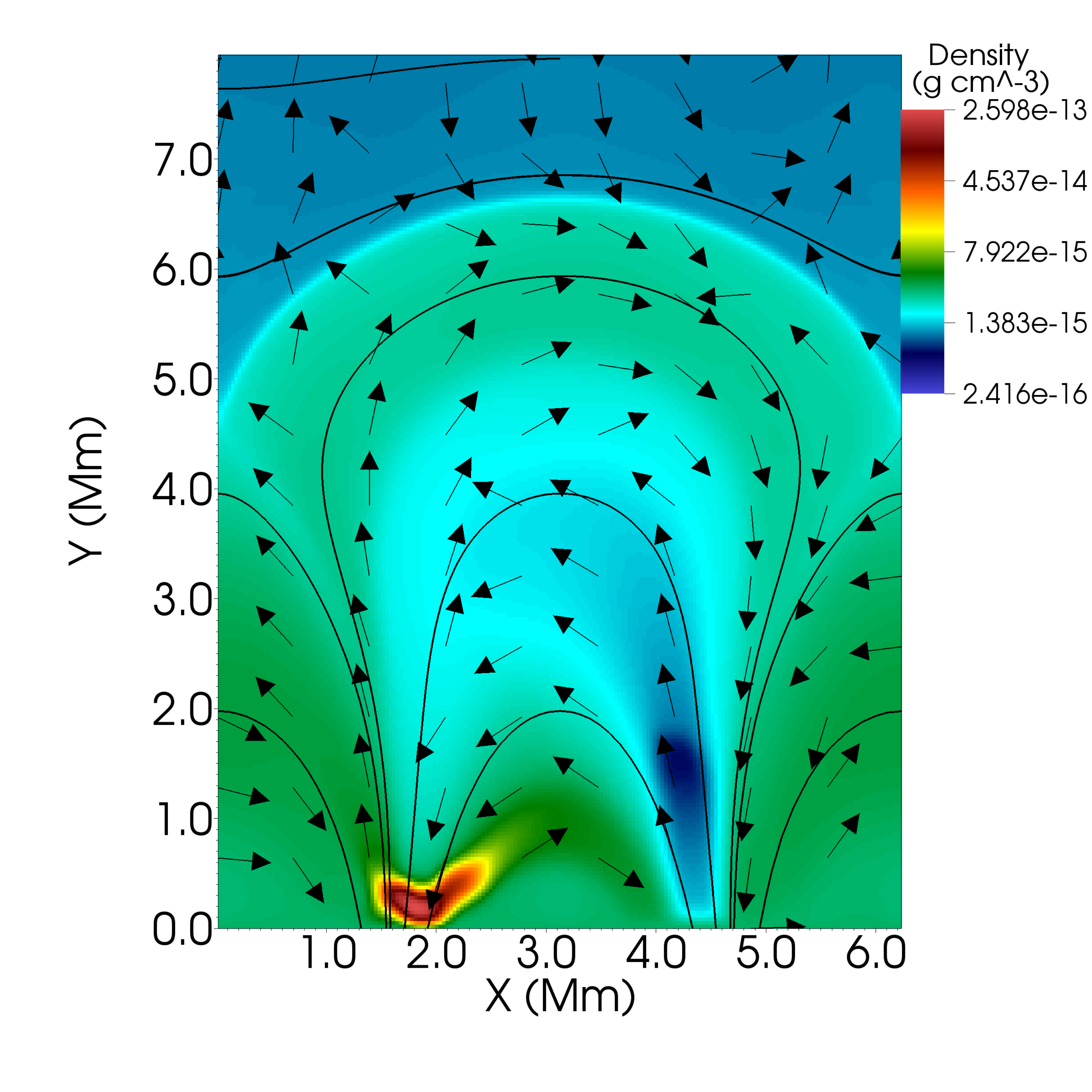}
    \caption{$t=83.01$ min}
    \label{fig:den_t1160}
\end{subfigure}
\centering
\begin{subfigure}{0.3\textwidth}
  \includegraphics[width=1\textwidth]{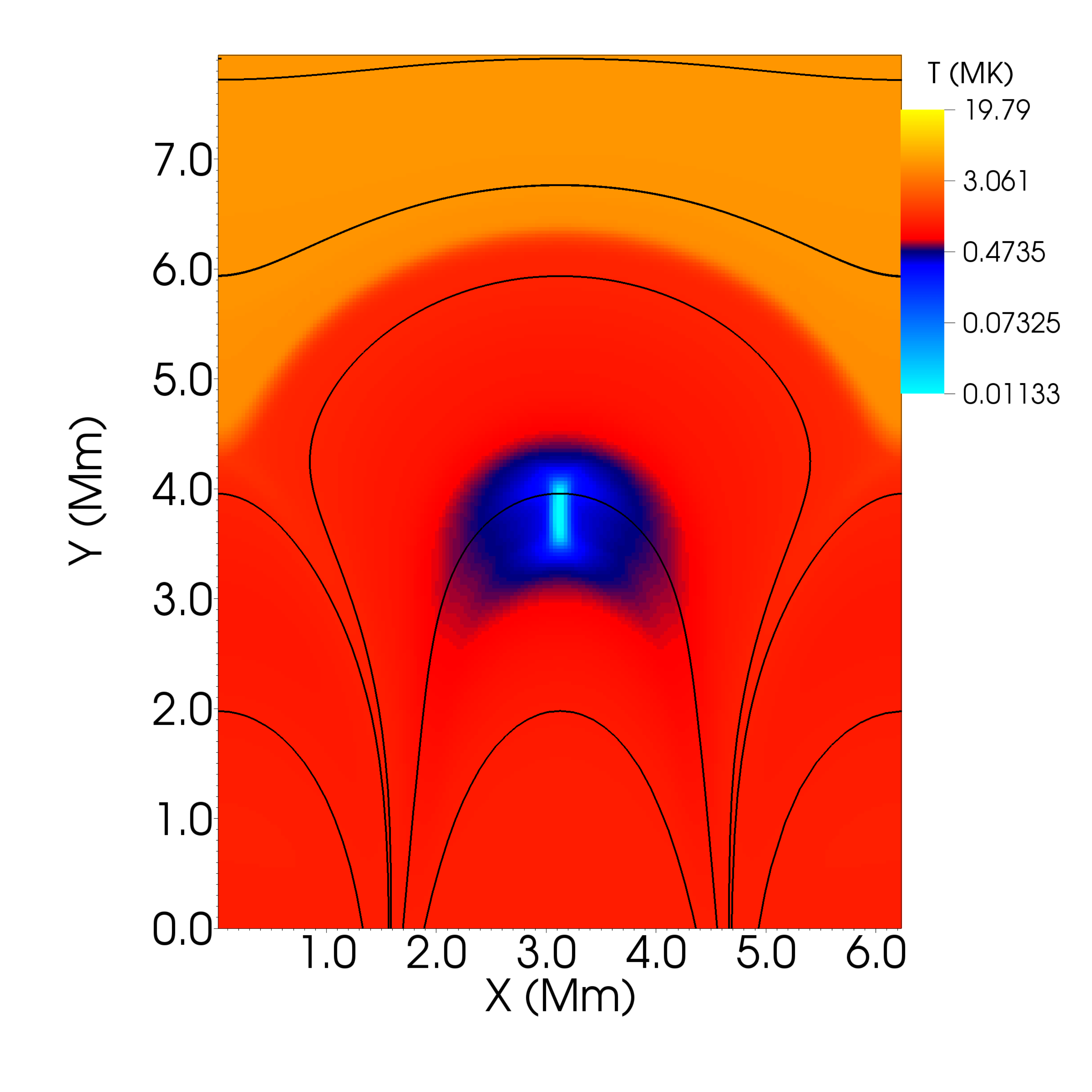}
   \caption{$t=64.83$ min}
   \label{fig:temp_t0906}
\end{subfigure}
\begin{subfigure}{0.3\textwidth}
    \includegraphics[width=1\textwidth]{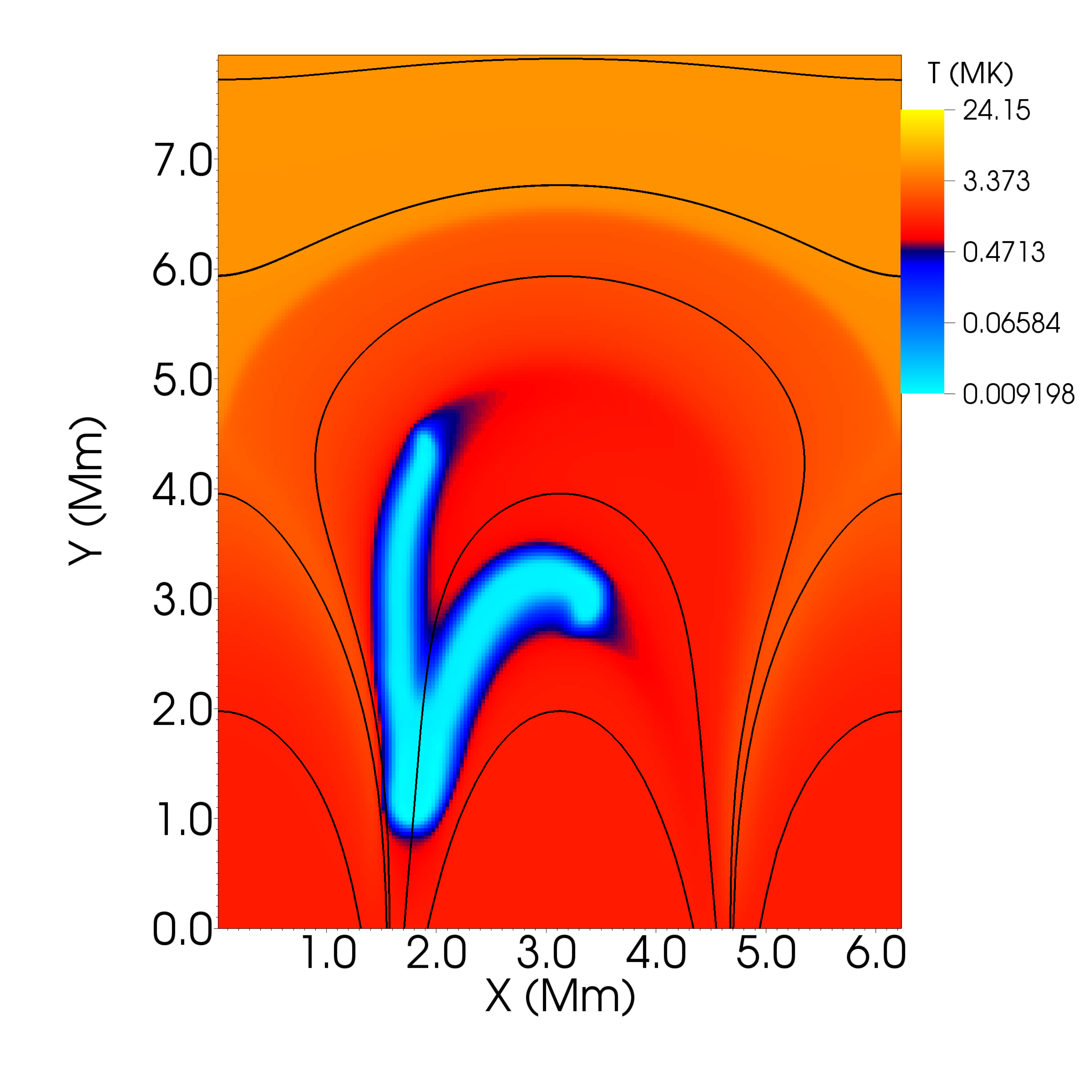}
    \caption{$t=80.10$ min}
    \label{fig:temp_t1120}
\end{subfigure}
\begin{subfigure}{0.3\textwidth}
    \includegraphics[width=1\textwidth]{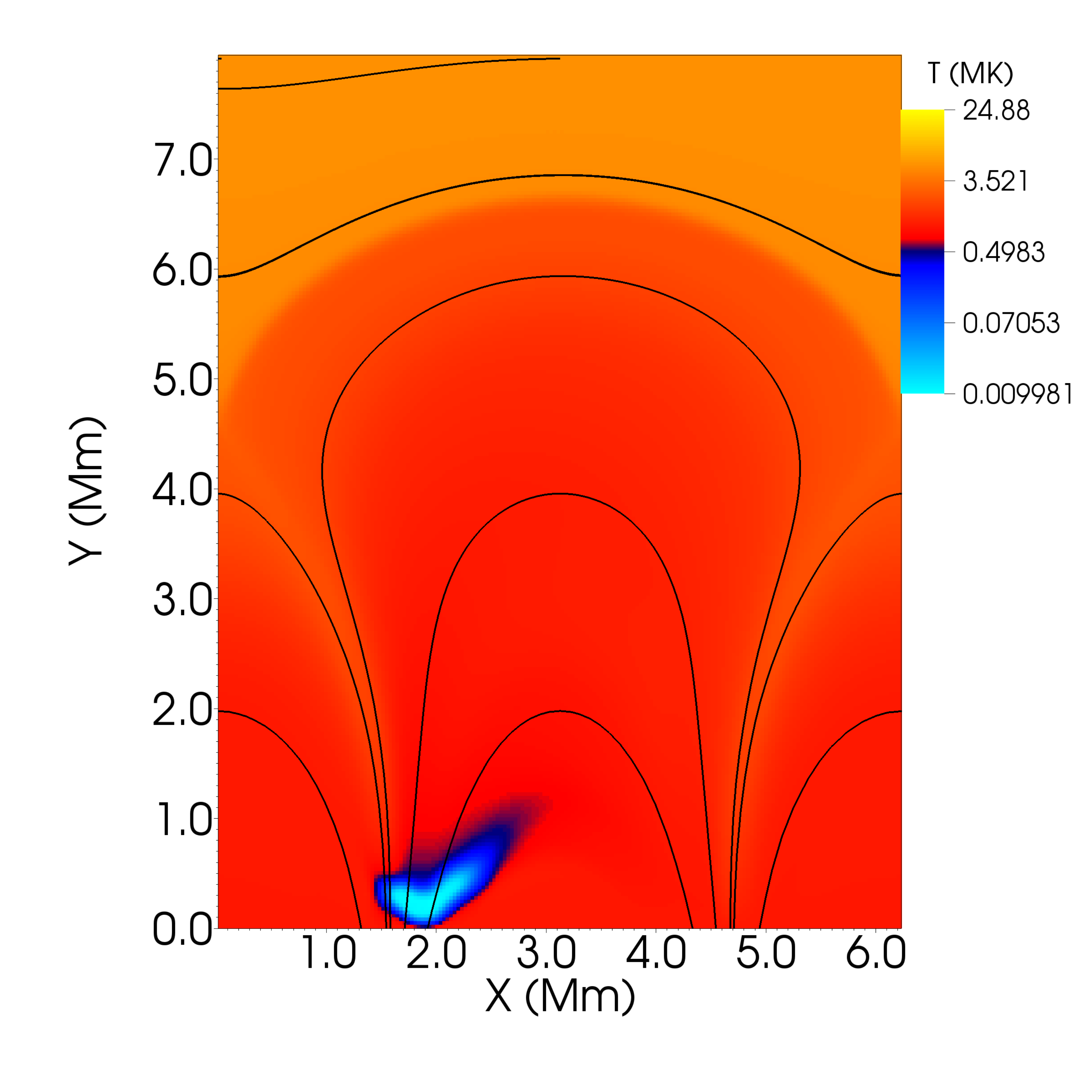}
    \caption{$t=83.01$ min}
    \label{fig:temp_t1160}
\end{subfigure}
\begin{subfigure}{0.3\textwidth}
  \includegraphics[width=1\textwidth]{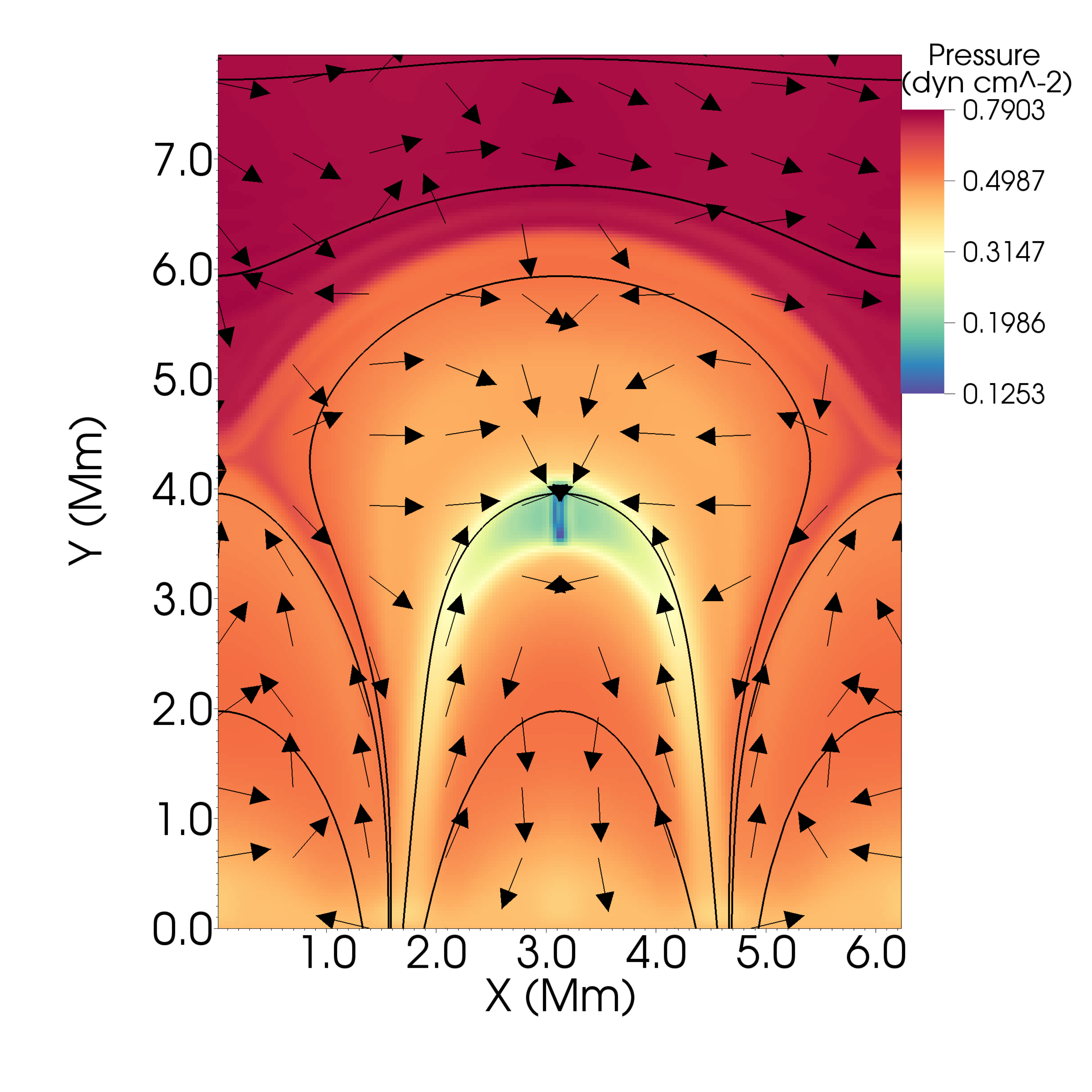}
   \caption{$t=64.83$ min}
   \label{fig:p_t0906}
\end{subfigure}
\begin{subfigure}{0.3\textwidth}
    \includegraphics[width=1\textwidth]{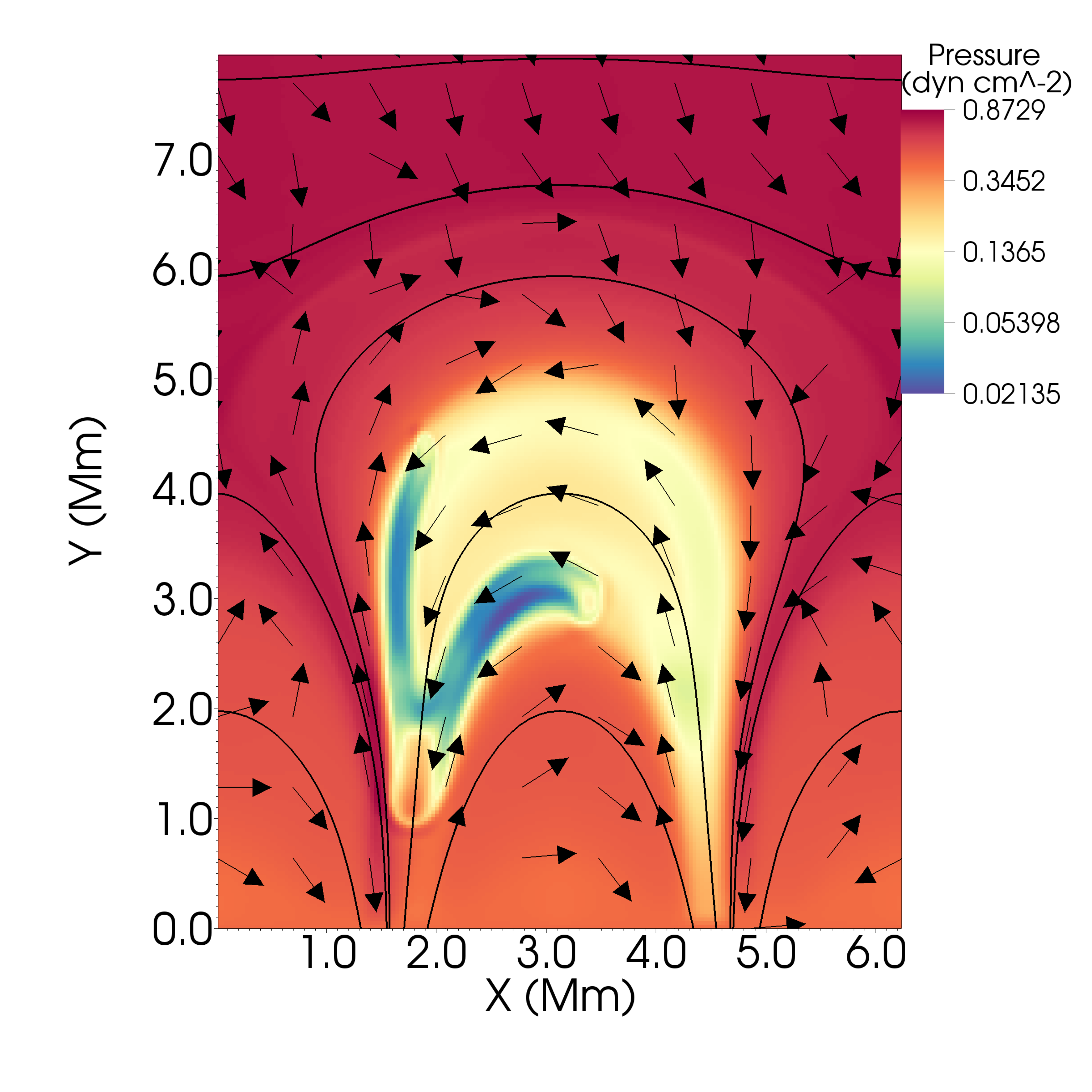}
    \caption{$t=80.10$ min}
    \label{fig:p_t1120}
\end{subfigure}
\begin{subfigure}{0.3\textwidth}
    \includegraphics[width=1\textwidth]{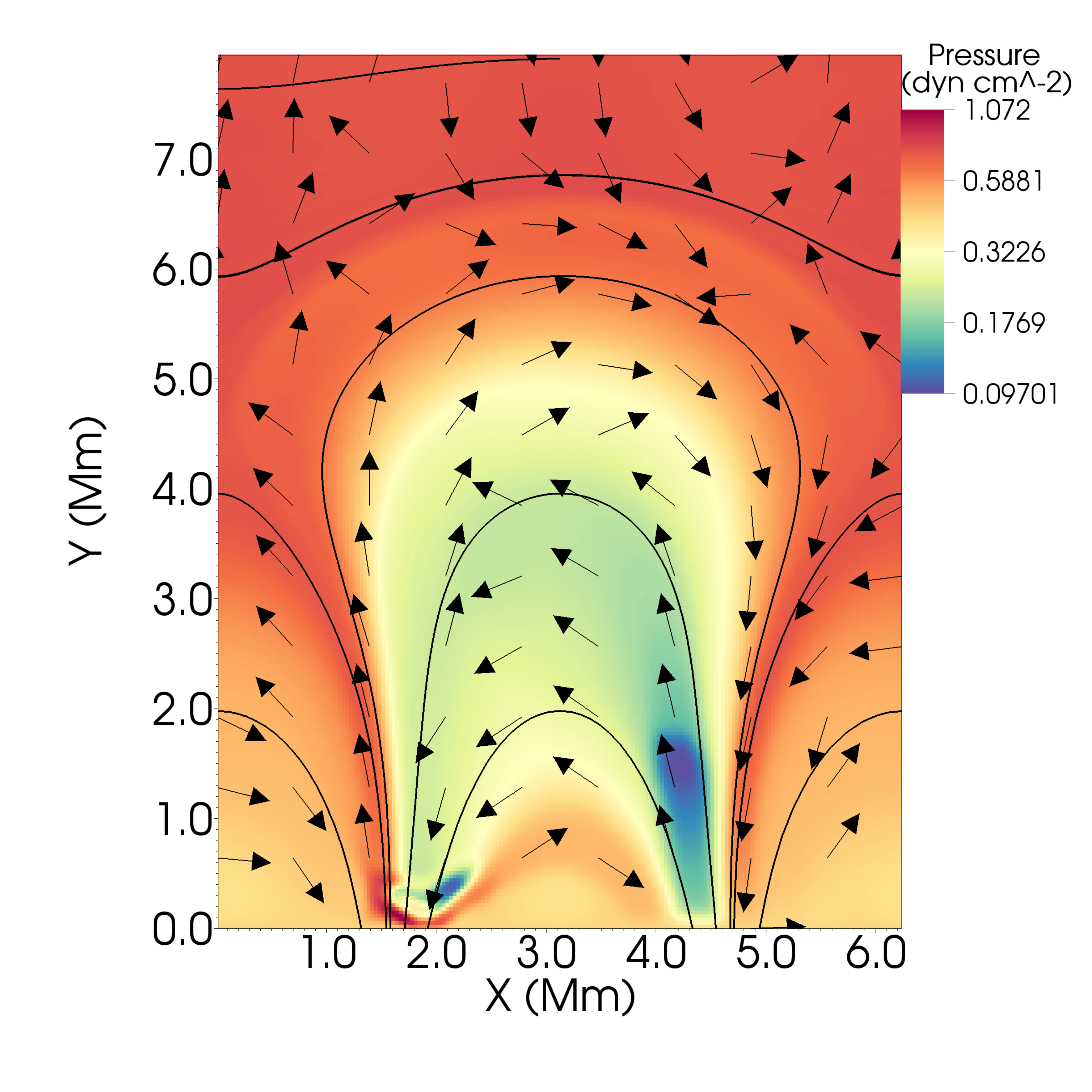}
    \caption{$t=83.01$ min}
    \label{fig:p_t1160}
\end{subfigure}
\caption{Density (top panel), temperature (middle panel), and gas pressure (bottom panel) in different stages when the cool ($\sim 10^4$ K), and condensed ($\sim 10^{-13}$ g cm$^{-3}$) coronal rain forms. The location of three separated density blobs is marked by the white boxes in (b). Overplotted black solid curves represent the magnetic field lines, and the arrows in the bottom panels represent the velocity field, $\lbrace v_x, v_y\rbrace$ in the $x-y$ plane.
} 
\label{fig:coronal-rain}
\end{figure*}

\begin{figure*}[hbt!]
\centering
\begin{subfigure}{0.33\textwidth}
  \includegraphics[width=1\textwidth]{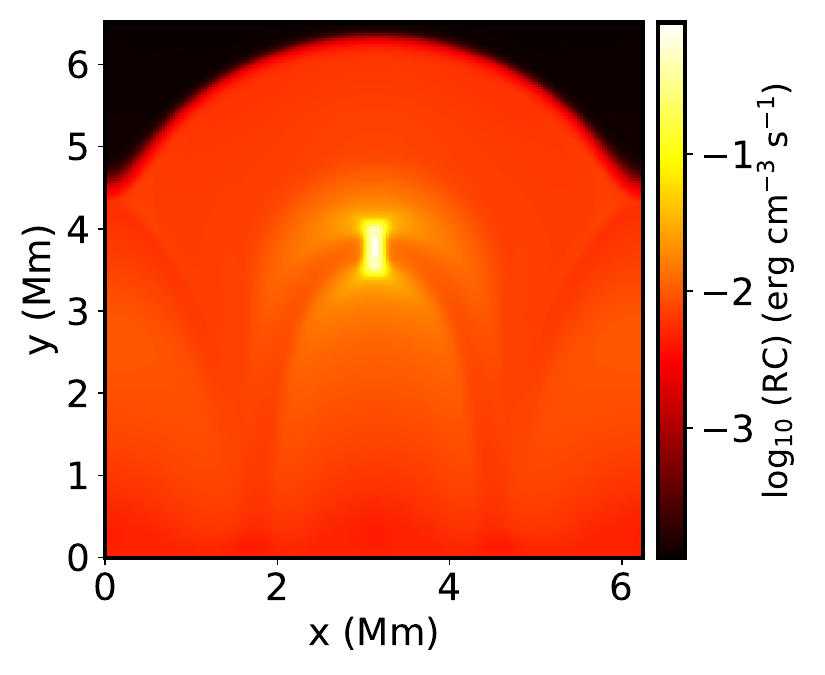}
   \caption{$t=64.83$ min}
   \label{fig:rc_t0906}
\end{subfigure}
\begin{subfigure}{0.33\textwidth}
    \includegraphics[width=1\textwidth]{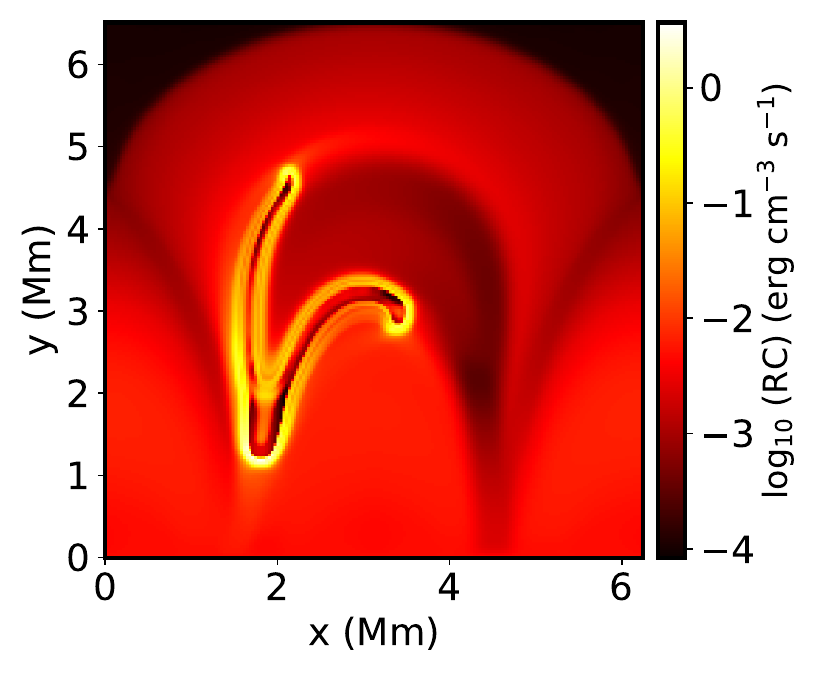}
    \caption{$t=80.1$ min}
    \label{fig:rc_t1120}
\end{subfigure}
\begin{subfigure}{0.33\textwidth}
    \includegraphics[width=1\textwidth]{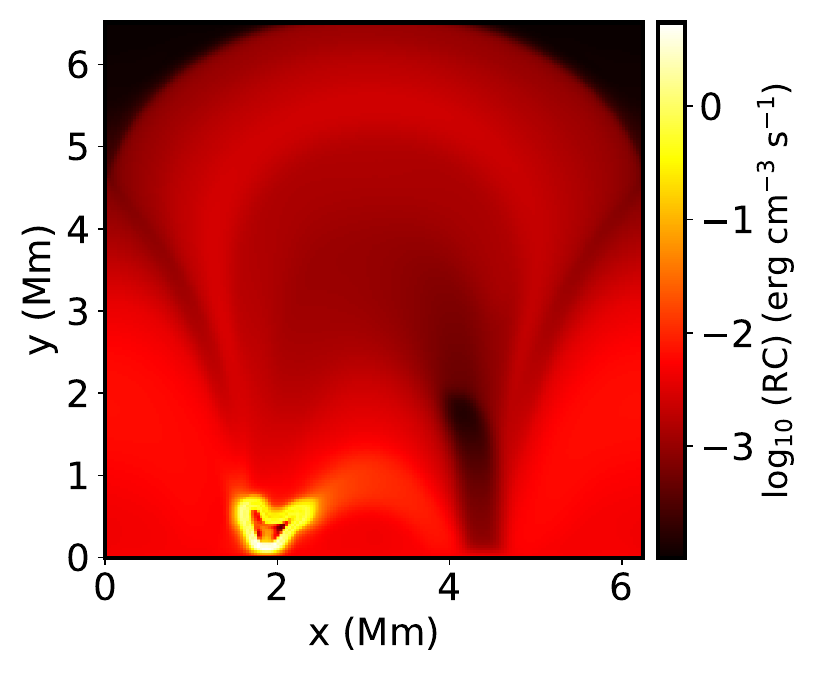}
    \caption{$t=83.01$ min}
    \label{fig:rc_t1160}
\end{subfigure}
\caption{Optically thin radiative cooling (RC) for the same time snapshots as in Figure \ref{fig:coronal-rain}.} 
\label{fig:rc}
\end{figure*}

\begin{figure}[hbt!]
    \centering
    \begin{subfigure}{0.5\textwidth}
    \includegraphics[width=1\textwidth]{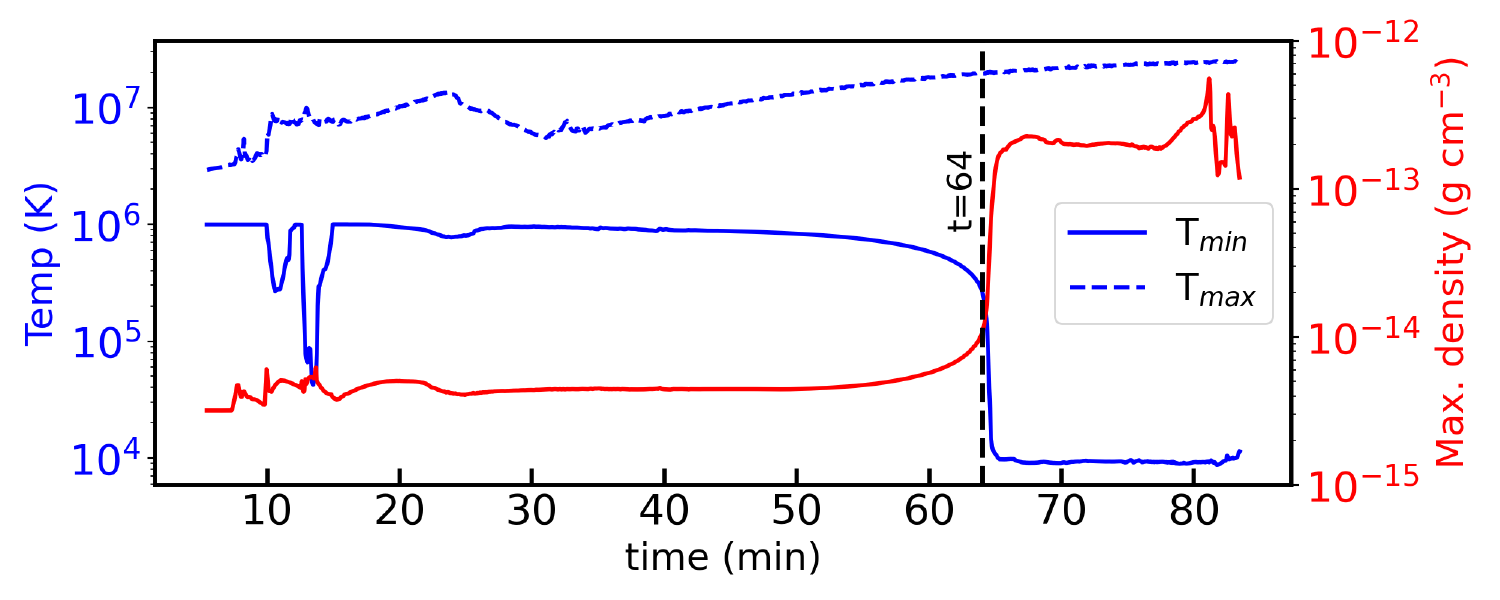}
    \caption{}
    \label{fig:den_temp_ts}
\end{subfigure}
\begin{subfigure}{0.36\textwidth}
    \includegraphics[width=1\textwidth]{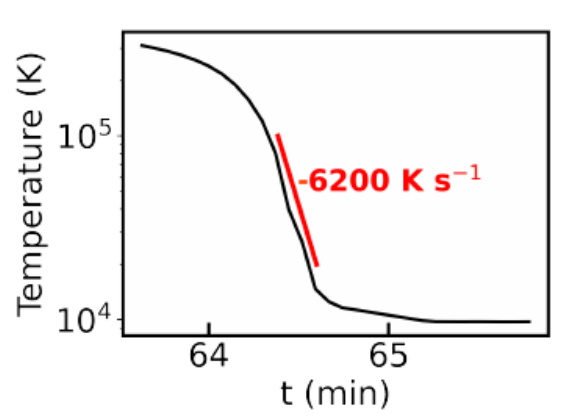}
    \caption{}
    \label{fig:coolin_rate}
\end{subfigure}
\caption{(a) Time series of instantaneous maximal and minimal temperature, and peak density. (b) represents the temporal variation of instantaneous minimum temperature highlighting the catastrophic cooling phase, where the maximum cooling rate is estimated by the slope $\approx -6200$ K s$^{-1}$ marked by the red line.} 
\end{figure}

\begin{figure}[hbt!]
    \centering
    \includegraphics[width=0.4\textwidth]{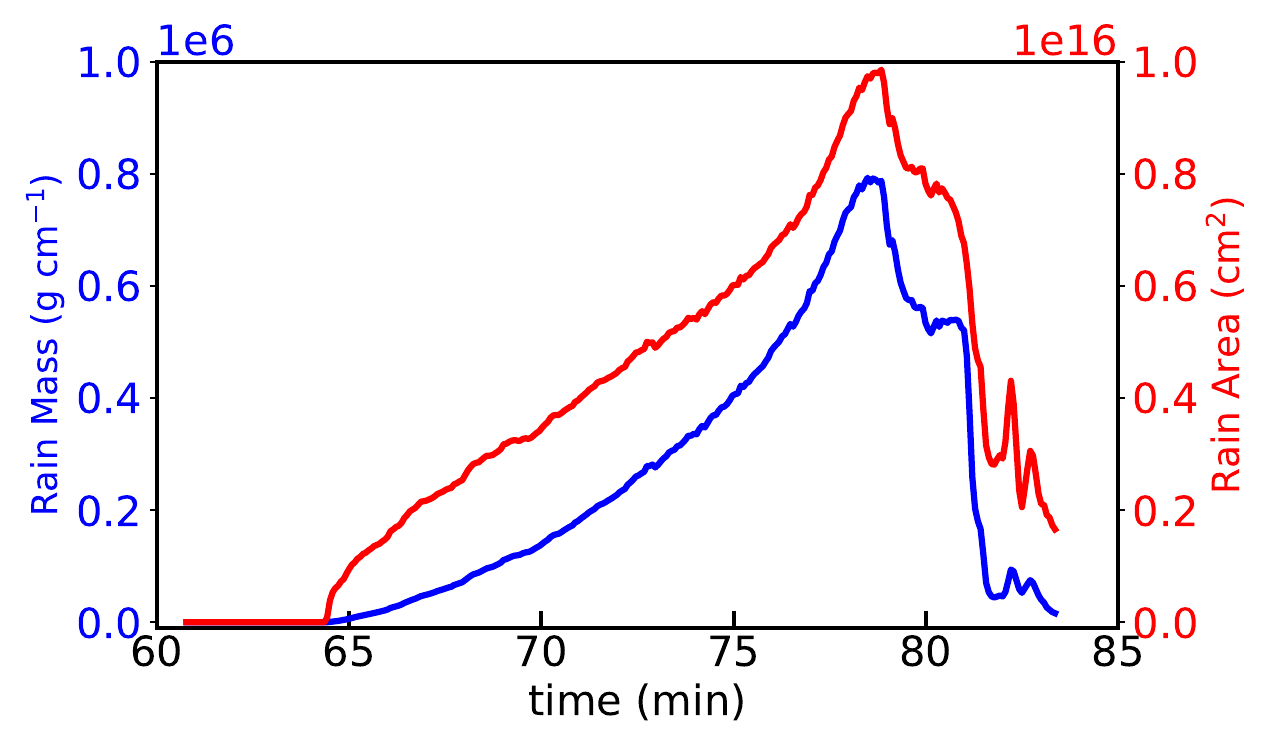}
    \caption{Variation of coronal rain mass and its area with time.}
    \label{fig:rain_mass_area}
\end{figure}

\begin{figure}[hbt!]
    \centering
    \begin{subfigure}{0.7\columnwidth}
    \includegraphics[width=1\textwidth]{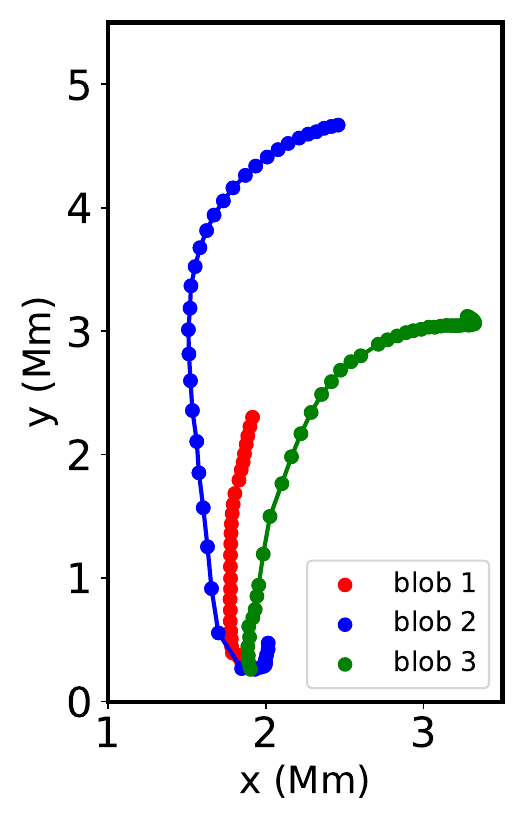}
    \caption{}
    \label{fig:track}
\end{subfigure}
\begin{subfigure}{0.8\columnwidth}
    \includegraphics[width=1\textwidth]{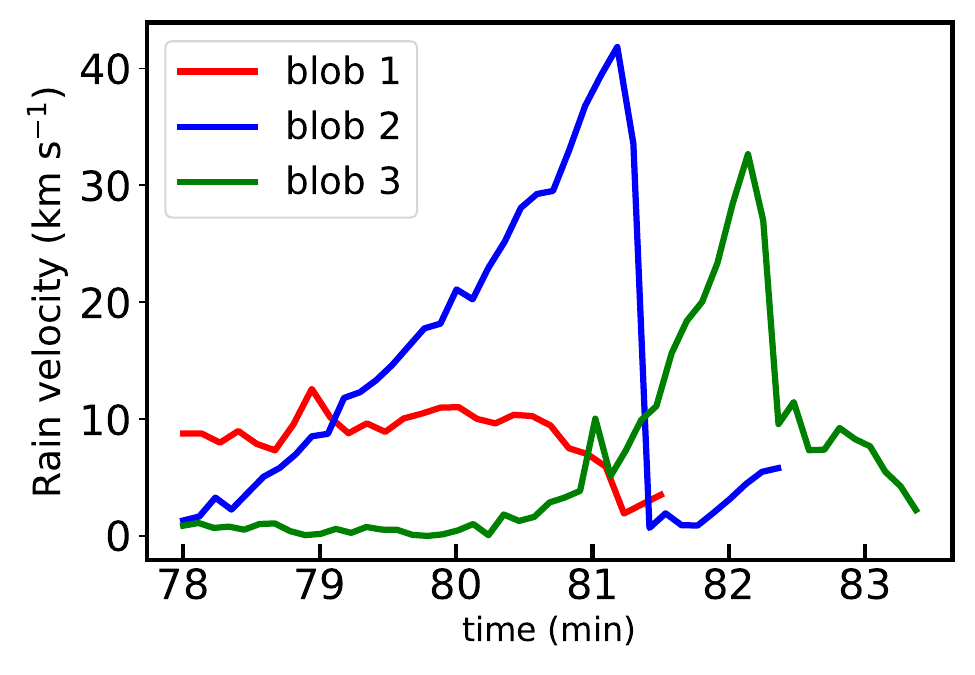}
    \caption{}
    \label{fig:rain_v}
\end{subfigure}
\caption{(a) shows the trajectories of the individual rain blobs on $x-y$ plane. (b) shows the instantaneous vertical velocities of the individual falling rain clumps.} 
\end{figure}

\subsection{Development of thermal instability}\label{sec:thermal instability}
As a consequence of post-eruption of the MFRs, the loop footpoints are heated due to the combined effect of (field-aligned) thermal conduction, steady background heat (which is based on the initial density and temperature), and radiative cooling. This leads to the evaporation of plasma materials along the field lines. To investigate the flows of plasma materials, we estimate the velocity along the projection of the magnetic field lines on the vertical $XY$ plane, namely, $\displaystyle{v_{fl} = \text{sign}(v_y) \ \frac{|v_x B_x + v_y B_y|}{\sqrt{B_x^2 + B_y^2}}}$, as shown in Figure \ref{fig:vfl}. The eruptions with several MFRs that escape upwards are followed by the flow of plasma materials along the field lines from the loop foot points to the central loop top. These materials accumulate at the loop top resulting in a gradual density enhancement. The radiative cooling rate in the optically thin corona, $H_{rc} \sim \rho^2 \Lambda(T)$, has derivatives  $\displaystyle{\frac{\partial\, \text{log}\,H_{rc}}{\partial\, \text{log}\,\rho} =2}$, and $\displaystyle{\frac{\partial\, \text{log}\,H_{rc}}{\partial \,\text{log}\,T} \approx -0.4}$ at $T \approx 1$ MK for the $\Lambda(T)$ cooling curve we use in this model. The temperature at the loop top location $(x, y) = (3.14, 4)$ Mm remains $\sim 1$ MK between $t\approx 20$ to $40$ min (see Figure \ref{fig:flare-duration}, top panel); this implies that the radiative cooling rate at $T \approx 1$ MK is more sensitive to the density perturbations than to those of temperature. The gradual mass accumulation at the loop top at $x=3.14$ and around $y=4$ Mm leads to an extended phase where a thermal instability slowly sets in and causes a local condensation in the post-flare loop top. We follow the temporal evolution of temperature at four different locations at $x=1, 2, 3.14$, and $4.7$~Mm all taken at the same height namely $y=4$ Mm, as shown in Figure \ref{fig:flare-duration} (top panel). The temperature at those locations can reach values up to several million degrees during the eruption phase. The maximum temperature at $(x,y) = (3.14, 4)$ Mm rises up to $\approx 7.6$ MK at the CS location beneath the MFR during the second central eruption. After the second MFR eruption from the central region, there is a secondary eruption from the side boundaries (as can be seen in the animations associated with Figure \ref{fig:den_early}), which leaves the simulation domain from the top boundary at $\approx 38$ min. Indeed, about half an hour into our simulation, no more MFRs erupt and the magnetic topology is seen to have changed into an arcade configuration with an overlying mostly horizontal field. The temperature evolution at the different selected $x$-values for $y=4$~Mm show minor thermal fluctuations around a typical coronal $1$~MK degree for $t=25$~min onward (see top panel of Figure \ref{fig:flare-duration}). These fluctuations are post-eruption effects. These perturbations settle down at $\approx 42$ min. The temperature at the loop top location $(x,y) = (3.14, 4)$ Mm gradually starts falling down right after that (as shown by the green curve in the top panel of Figure \ref{fig:flare-duration}). This signals the gradual development of thermal instability at the post-eruption loop top. The temperature drop occurs as the cooling rate, which is a combination of optically-thin radiation and field-aligned thermal conduction dominates over the (steady but spatially varying) background heating given by equation (\ref{eq:Hbgr}). The temperature evolution at this $(x,y) = (3.14, 4)$~Mm location in between $t=42$ and $50$~min represents the (nearly) linear decay phase of temperature as shown by the bottom panel of Figure \ref{fig:flare-duration}. The temperature decay rate within this period is estimated to be $\approx 156$ K s$^{-1}$ by the slope marked by the red dashed line. The steady background heating rate at the loop top location is $\approx 4.33 \times 10^{-3}$ erg cm$^{-3}$ s$^{-1}$. Figure \ref{fig:heating-cooling-misbalance} shows that the net heating rate due to conduction and background heat dominates over the radiative cooling rate at the loop top location at the earlier stage of post eruption (between $t \approx 20$ and $40$~min). Later, radiative cooling gradually starts to dominate over the heating from around $t=40$ min onwards, and at $t=64$ min the radiative cooling rate increases sharply, when a catastrophic thermal runaway process starts. 
This leads to a pressure drop at the loop top location compared to its surroundings (see Figure \ref{fig:p_t0906}), which drives pressure gradient flows into that location, and leads to a rapid density enhancement and drop in temperature (shown by the right most vertical dashed line in the top panel of Fig. \ref{fig:flare-duration}).


\subsection{Evolution of coronal rain}\label{coronal rain}
The location of the first condensation seed that appears in the central loop top is shown in Figure \ref{fig:den_t0906}. What happens next is shown visually in Figure~\ref{fig:coronal-rain}, which contains three rows, for density, temperature and pressure, and three columns, for $t=64.83$~min, $t=80.10$~min and $t=83.01$~min. Figures~\ref{fig:den_t0906} and \ref{fig:temp_t0906}, shows the maps of the density and temperature at $t=64.83$~min. The density is enhanced up to $\sim 10^{-13}$~g~cm$^{-3}$ which is around $100$ times more than the initial density ($\sim 10^{-15}$ g cm$^{-3}$), and the temperature drops to $\sim 10$ kK, which is around $100$ times lower than the initial temperature (1 MK). We see that this thermal runaway process initiates only at the magnetic loop top. The dips in the temperature curves in Figure \ref{fig:flare-duration} (top panel) for both the $(x, y)= (2, 4)$ and (3.14, 4) Mm locations (shown by the blue and green curves respectively) show what happens in the temperature evolution at the fixed points, and must be understood from the evolution as shown in Fig.~\ref{fig:den_t0906}-\ref{fig:den_t1160}. The condensation accumulates surrounding plasma, which is driven by the pressure gradient flows due to the pressure drop in the condensation site. The pressure is as shown in the bottom panels of Figure \ref{fig:coronal-rain}, where we notice the inflow of matter from higher to lower gas pressure regions towards the condensation sites. In Figure \ref{fig:den_t1120}, we see that the loop top condensation has split into three isolated density clumps (or `blobs') as marked by the white boxes. These clumps fall downward along the magnetic field lines due to the combination of gravity and the Lorentz force, and they finally merge together again as shown in Figure \ref{fig:den_t1160}, at $t=83.01$~min. These cool-condensed structures share similarities with thermodynamic properties of coronal rain in the solar atmosphere as observed by \cite[][and references therein]{2005:muller, 2015:patrick, 2016:scullion, 2019:mason, 2023:seray}, where the density and temperature of the rain clumps are reported between $\sim 10^{-14} - 10^{-13}$ g cm$^{-3}$, and $\sim 10^4-10^5$ K respectively.

From Figures \ref{fig:temp_t0906}-\ref{fig:temp_t1160}, we notice that the cool ($\sim 10^4$ K) material is spatially co-located with the condensed structures in Figures \ref{fig:den_t0906}-\ref{fig:den_t1160}. The corresponding radiative cooling losses are shown in Figure \ref{fig:rc}. The maximum heat loss rate due to (optically thin) radiation in Figure \ref{fig:rc_t0906} is $\approx 9.06 \times 10^{-1}$ erg cm$^{-3}$ s$^{-1}$ which dominates over the background heating of $\approx 4.33 \times 10^{-3}$ erg cm$^{-3}$ s$^{-1}$. The maximum cooling rate increases at later times to $3.65$ and $5.47$ erg cm$^{-3}$ s$^{-1}$, as shown in Figures \ref{fig:rc_t1120} and \ref{fig:rc_t1160}, respectively. The regions with high cooling rates correspond to the cool-condensed regions, as shown in the top and middle rows of Figure \ref{fig:coronal-rain}. Our 2.5D assumption influences the radiative cooling appearance, showing the condensations as dark structures with bright edges, akin to the prominence-corona-transition-region (PCTR) transition obtained in other recent 2.5D coronal rain simulations \citep{2022ApJ...926..216L}.

The instantaneous maximum and minimum temperature evolution within the entire simulation domain are shown in Figure \ref{fig:den_temp_ts}, covering the entire period from eruptions to the late condensation stages. The minimum temperature distribution has two strong dips between $\approx 10-15$ min. The first dip appears when the plasma blobs which are denser than the surrounding medium, are trapped inside secondary erupting flux ropes from the side boundaries, which also correspond to lower temperature than the background medium (see Figures \ref{fig:den_early}(c), and \ref{fig:den_early}(g)). The second dip at about 14 minutes appears due to a trapped plasma blob inside the rope from the second central eruption (see Figures \ref{fig:den_early}(d), and \ref{fig:den_early}(h)). The corresponding signatures are also evident in the red curve in Figure~\ref{fig:den_temp_ts}, which quantifies the instantaneous maximum density. The same figure shows later on a sharp rise in this peak density (from $\sim 10^{-15}-10^{-13}$ g cm$^{-3}$), and drop in minimum temperature from $\sim$ MK to $\sim 10$ kK at $\approx 64$ min. This signals the onset of the thermal runaway process to form the cool condensation at the loop top location $(x, y)=(3.14, 4)$ Mm. This catastrophic drop in (minimal) temperature happens between $\approx 64$ to 65 min, and the sudden temperature change during that short time interval is highlighted in Figure \ref{fig:coolin_rate}. The maximum rate of change of the temperature during this minute is $\approx 6.2$ kK s$^{-1}$. This catastrophic cooling rate is in good agreement with a 2.5D MHD simulation \citep{2021:wenzhi} of a post-flare coronal rain scenario, where they found the average catastrophic cooling rate to be 9 kK s$^{-1}$, and in fair agreement with observations for post-flare arcades \citep{2016:scullion}, where it is reported to be $22.7$ kK s$^{-1}$. The estimation of the cooling rates sometimes can be challenging in the observations, since it requires measurements of different spectral lines with different instruments and their responses, sampling different parts of the flare. Note that the details of the cooling curve prescription, $\Lambda(T)$, will also influence the condensation rates, as was shown in a pure coronal volume study which compared many popular cooling curves in \cite{2021A&A...655A..36H}.
The temporal evolution of the rain mass and its area are shown in Figure \ref{fig:rain_mass_area}. Here, we define the localized condensations as coronal rain clumps when the density of those regions is more than or equal to a density threshold value, which is 10 times the maximum density of the initial state. To calculate the total rain mass at an instantaneous time, we collect all the cells with densities above the threshold, then multiply the local densities with the area of the individual cells to obtain the mass per unit length of those individual cells, and finally aggregate the quantity of all those cells. The evolution of the rain mass and area increase sharply from the onset time of thermal runaway at $\approx 64$ min. This is also consistent with Figure \ref{fig:den_temp_ts} since the catastrophic temperature drop occurs at the same time. The rain mass and area reach maximum values at $\approx 78.5$ min. At about this time, the condensed region is separated into three individual density blobs (shown in Figure \ref{fig:den_t1120}). Thereafter, we see a decrease in rain mass and area. This is because the blobs start to coalesce while falling down, finally forming a single blob again (see Figure \ref{fig:den_t1160}). The decrease of the rain mass and area after 79 minutes is also influenced by the fact that the smaller blobs are more easily evaporated again, as they radiate energy in their PCTR and that they attain lower heights where the background heating is higher. To track the motion of individual rain blobs, we use an automated method using binary masking from a \texttt{python} based module, \texttt{OpenCV}\footnote{more information can be found at \href{https://docs.opencv.org/3.4/d3/dc0/group__imgproc__shape.html}{https://opencv.org/}} to locate the instantaneous centroid locations of the density blobs. We start the tracking when it first separates into three blobs (around $t\approx 78$ min), and continue until the end of the simulation when they merge together upon reaching the bottom ($t \approx 83.37$ min). Our `blob 1' first falls down at the bottom, followed by `blob 2' coalescing with `blob 1', and finally the `blob 3' falls down and merges to form a single blob. The trajectories of the individual blobs are shown in Figure \ref{fig:track}, where the falling blobs follow trajectories (nearly) along magnetic field lines (see the animation associated with Figure \ref{fig:den_early}). Instantaneous vertical velocities of the individual blobs are calculated based on the instantaneous distance traversed along the vertical direction ($y$-direction) by the blobs between two consecutive simulation frames, taken at a temporal cadence of 4.29 s. The evolution of the falling rain velocity is shown in Figure \ref{fig:rain_v}, where we notice that the velocity of the rain clumps is $\approx 1-40$ km s$^{-1}$. It increases as they accelerate while falling downward due to the combined effect of gravity, gas pressure gradients, and LF. This is clearly noticeable, especially for `blob 2' and `blob 3' (in Figure \ref{fig:rain_v}), where they obtain the maximum velocities when they reach the bottom just before coalescing with other. They are decelerated suddenly, which is due to our lower boundary condition and pressure gradient. The velocity of the downward falling rain clumps in our study is in good agreement with other simulation models. Examples include \cite{2015:fang}, where the maximum velocity of the blobs in a rain shower are $\approx 60$ km s$^{-1}$, with most of the rain blobs having velocities  $\approx 30$ km s$^{-1}$, and \cite{2022ApJ...926..216L}, where the velocity varies from a few to 120 km s$^{-1}$. It is also in reasonable agreement with the the observations of \cite{2015:patrick} and \cite{2019:mason}, who reported velocities of the rain clumps in the range $\approx 50 - 100$ km s$^{-1}$.

\section{Summary and Discussion} \label{sec:summary-discussion}

In this work, we use a 2.5D resistive MHD simulation incorporating non-adiabatic effects of optically thin radiative cooling, field-aligned thermal conduction, and steady (but spatially varying) background heating to explore a series of flux rope eruptions, followed by gradual development of thermal instability, ultimately leading to a catastrophic cooling phase in the solar corona. The initial magnetic field configuration used  in this simulation is motivated from \cite{2016:sanjay}, where they study the magnetic and dynamical properties of an erupting flux rope in an incompressible medium. Here, we use the similar initial magnetic field prescription, but extend the study using a gravitationally stratified solar atmosphere in a compressible medium, and the multi-thermal behaviour in the solar corona. 

The initial magnetic field is non-force-free, and its strength depends on three parameters, $\sigma$,  $\gamma$ (which is related to the shearing angle), and the field strength $B_0$. These initial forces instantly induce flows that drive the plasma evolution. We showed how this imbalance transiently evolves to a more force-balanced, energetically settled semi-equilibrium stage after 5.44 minutes, which we use as a reference point for the rest of the simulation. The Lorentz force-driven flows cause convergence of plasma along the central axis and trigger magnetic reconnection leading to flux rope formation and eruptions.
The shearing angle of the magnetic arcades is also one of the important parameters for the formation of the flux ropes, and our simulation shows clear successive MFR formations with a shearing angle of $\gamma = 72^\circ.5$. To investigate the effect of a lower $\gamma$, we also ran the simulation up to $\approx 90$ minutes with $\gamma = 25^\circ$ keeping all the other parameters the same (not shown in our figures), and notice that there is no formation (and eruption) of flux ropes. Successive MFR eruptions are observed in a multiple flux rope system in a solar flare \citep{2018:hou}, where two consecutive ``double-decker flux rope'' eruptions happen within an interval of five minutes due to strong shearing motions. This is very similar with our findings, where the interval between two (central) eruptions in our study is $\approx 4.9$ minutes. Multiple MFR eruptions are also found in other simulations, such as in three-arcade configurations by \cite{2006:ding} in a resistive medium, and \cite{2009:soenen} in an ideal MHD regime. These studies show homologous eruptions of multiple flux ropes from the same sites, which agrees with our findings where the two MFRs are ejected from the central arcade region.
However, the main focus of our study is to investigate the multiple eruption of flux ropes and the multi-thermal behaviour in a localized domain in corona. This present setup is not motivated to portray a global MHD model, where it fits in the larger picture of CME evolutions, e.g. due to breakout scenario, and how the large-scale magnetic topology (at further out radii, i.e. in the top part of our simulation domain) evolves in solar corona. We believe that the use of periodic boundary condition at the side be adopted meaningfully and in fact a standard procedure for a local-box simulation. Examples include, \cite{2016:sanjay}, where they have studied the eruption, and \cite{2022:robinson}, where the formation of a pre-flare flux rope is studied in periodic side boundaries. In our simulation, the magnetic fields above the central arcade switch to a primary horizontal orientation when the eruptive flux ropes expand laterally, and it is of course mediated by the choice of periodic boundary condition at sides. We witness a flipping of the horizontal magnetic field ($B_x$ component) at the top part of the domain after each flux rope eruption. This is because the magnetic field (and mainly its horizontal component as introduced by the flux rope formations) due to each eruption dominates over the remnant magnetic field of the previous eruption. However, we perform an additional simulation with a modified boundary condition, where we use a three-arcade system, and by extending the lateral and vertical domain of our simulation box (see Appendix \ref{appendix}). Changing into this setup also gives the similar evolution: multiple eruption of flux ropes from an initial non-force-free setup, and the late stages of the post-flare arcade system that shows the formation of coronal condensation in a localized domain in corona. It does change (obviously) the way the eruptions proceed at the top part of the domain, but qualitatively the evolution are common in both cases simulated. We also notice that the central flux ropes expand laterally while eruption, and create horizontal field overlying the arcades. We admit that the details of the magnetic field topology in the top part of our domain, which drastically changes as successive flux ropes form and pass through, is not fully realistic in neither of the two simulated scenarios. We understand the importance of a more realistic setup which needs to be handled better in future, where e.g. the actual expansion radially is allowed, and the neigbouring magnetic flux evolutions are incorporated, such as when doing a full spherical model, and portray a large-scale (global) magnetic field of solar corona. Nevertheless, our main findings (multiple eruptions and a post-flare multi-thermal evolution) is not affected by this. The temperature evolution at the CSs underneath the MFRs during the second central eruptions at $(x,y) = (3.14, 4)$ Mm goes up to $\approx 7$ MK (from an initial 1 MK temperature). This high temperature would show clear brightenings in the soft X-ray band, as relevant for flares. 

The development of the thermal instability is noticed to start after eruption of the flux ropes (see Figure \ref{fig:flare-duration}).
 It first leads to a linear decay of temperature followed by a later catastrophic temperature drop and formation of a condensation that evolves, splits and falls down due to gravity. The cool material basically falls along the magnetic field lines of the loop. The thermodynamic behaviour of these cool-condensed structures is in good agreement with coronal rain observed in the solar atmosphere. We notice the formation of the condensation in the post-flare loop after $\sim 30$ min of the last flux rope eruption. This timescale is in agreement with the observation for a C8.2-class post-flare coronal rain, where the time difference between the flare ribbon formation, and the first appearance of coronal rain in the post-flare loops is $\approx 26$ min \citep{2016:scullion}. The observational evidence for coronal rain formation at the loop top regions that fall downwards along the magnetic field lines is similar to what we find in our simulation. The condensation time scale is also in a reasonable agreement with an $X2.1$ class post-flare coronal rain observation, where the time gap between the peak phase of the flare, and the first appearance of the condensation at the loop top is $\approx 50$ min \citep{2024:Brooks}. This indicates that our model is in close correspondence with actual post-flare coronal rain scenarios. The clumpy structure of the plasma blobs in our simulation is very much a feature of flare-driven rain, more so than in the quiescent (active region loop) rain, because the densities are much larger in the flare case than in quiescent rain. The density and temperature structuring of the rain, forming a coronal sheath around the cooler and denser cores of the rain clumps we found in our study is one of the unique aspects of the observations reported in \cite{2016:scullion}. They detected the source of the formation of the rain first in the hotter, and progressively later in the cooler Extreme-ultraviolet (EUV) channels. 

The role of resistivity $\eta$ in our model is an important aspect in the formation of MFRs and the CSs they involve, but also regulates current-driven Ohmic heating in all later phases where coronal condensations can form. The uniform resistivity we use in this model is $\eta = 5 \times 10^{-5}$ (in dimensionless units). To explore the effect of enhanced resistivity in the condensation process, we have also run the experiment with a higher value, $\eta = 5 \times 10^{-4}$ (in dimensionless units), keeping all the other parameters unchanged. 
We do not see any occurrence of a catastrophic cooling phase in the system till $90$~min. A more detailed parametric survey of $\gamma$, $\eta$, $\sigma$, and $B_0$ for this model may be interesting to study in the future to understand the dependence of the eruption and condensation mechanisms on those parameters. 

The maximum temperature in the upper half of the simulation domain reaches high values of up to $\approx 25$ MK (see Figure \ref{fig:den_temp_ts}). This is influenced by the fact that the successive MFRs also tend to evacuate this upper region, causing it to be heated more easily. The heating occurs due to our steady background heat prescription, which is based on the initial density and temperature values. The mass outflow from the top boundary due to the eruptions, and the prescription of no inflow through it leads to a density drop in that upper region, and therefore the background heat dominates over the cooling, leading to a high temperature. The use of the background heat prescription in this model is motivated by the fact that it is exactly compensated by the radiative cooling term in the initial state (there is no cooling due to thermal conduction at the initial stage because of our initial isothermal condition). Future extensions of this work should consider the detailed coupling to the lower chromosphere and photospheric regions. A localized heating prescription in the loop foot points can also trigger thermal non-equilibrium (TNE) to promote and regulate the appearence of coronal condensations, as proposed by \cite{1991:antiochos}. A recent study by \cite{2022ApJ...926..216L} demonstrates the formation of coronal rain showers due to turbulent heating at the chromospheric foot points. Advancement of our model by extending the spatial domain from the chromosphere to the corona and incorporating localized heating at the loop foot points can be interesting to carry out, which we plan to explore in future simulations. This may produce a quasi-cyclic formation of a larger number of rain clumps, or rain showers.

To conclude, our current model demonstrates the ability to reproduce the formation of post-flare coronal rain, where multiple flux rope eruptions happen from the same sites in the form of homologous ejections. It also sets direction for a better understanding of the the multi-thermal ($\sim$ 10 kK - 10 MK) behaviour in the solar corona, which is important to reveal the broader aspect of coronal heating.

\begin{acknowledgements}
Data visualization and analysis were performed using \href{https://visit-dav.github.io/visit-website/index.html}{Visit} and \href{https://yt-project.org/}{yt-project}. The authors thank the anonymous referee for the constructive suggestions, and also express their thanks to F. Moreno-Insertis, Eamon Scullion, Eric Priest, and Manuel Luna for the useful discussions and suggestions, that improved the manuscript considerably. SS and AP acknowledge support by the European Research Council through the Synergy Grant \#810218 (``The Whole Sun”, ERC-2018-SyG). SS and RK acknowledge support by the C1 project TRACESpace funded by KU Leuven, the ERC grant agreement 833251 Prominent ERC-ADG 2018, and FWO grant G0B4521N. RK acknowledges ISSI in Bern, Team project \#545. AP would like to acknowledge the support from the Research Council of Norway through its Centres of Excellence scheme, project number 262622. The computational resources and services used in this work were provided by the VSC (Flemish Supercomputer Center), funded by the Research Foundation Flanders (FWO) and the Flemish Government - department EWI; La Palma supercomputer, provided by Instituto de Astrof{\'i}sica de Canarias (IAC).  
\end{acknowledgements}

\bibliographystyle{aa} 
 \bibliography{ref}

\begin{appendix}
\section{Evolution of the system with modified setup}\label{appendix}

\begin{figure*}[hbt!]
\centering
\includegraphics[width=0.8\textwidth]{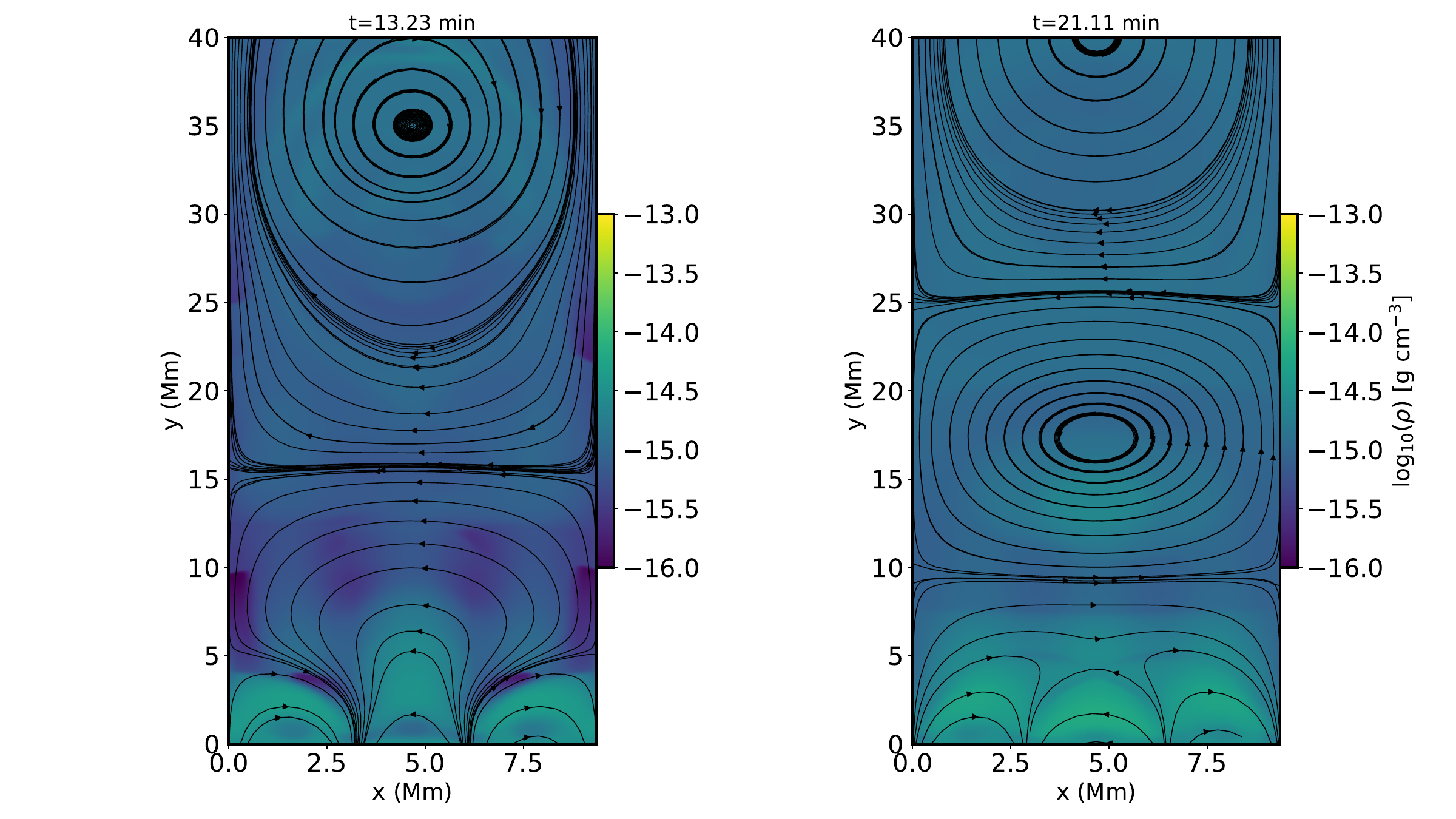}
\includegraphics[width=0.8\textwidth]{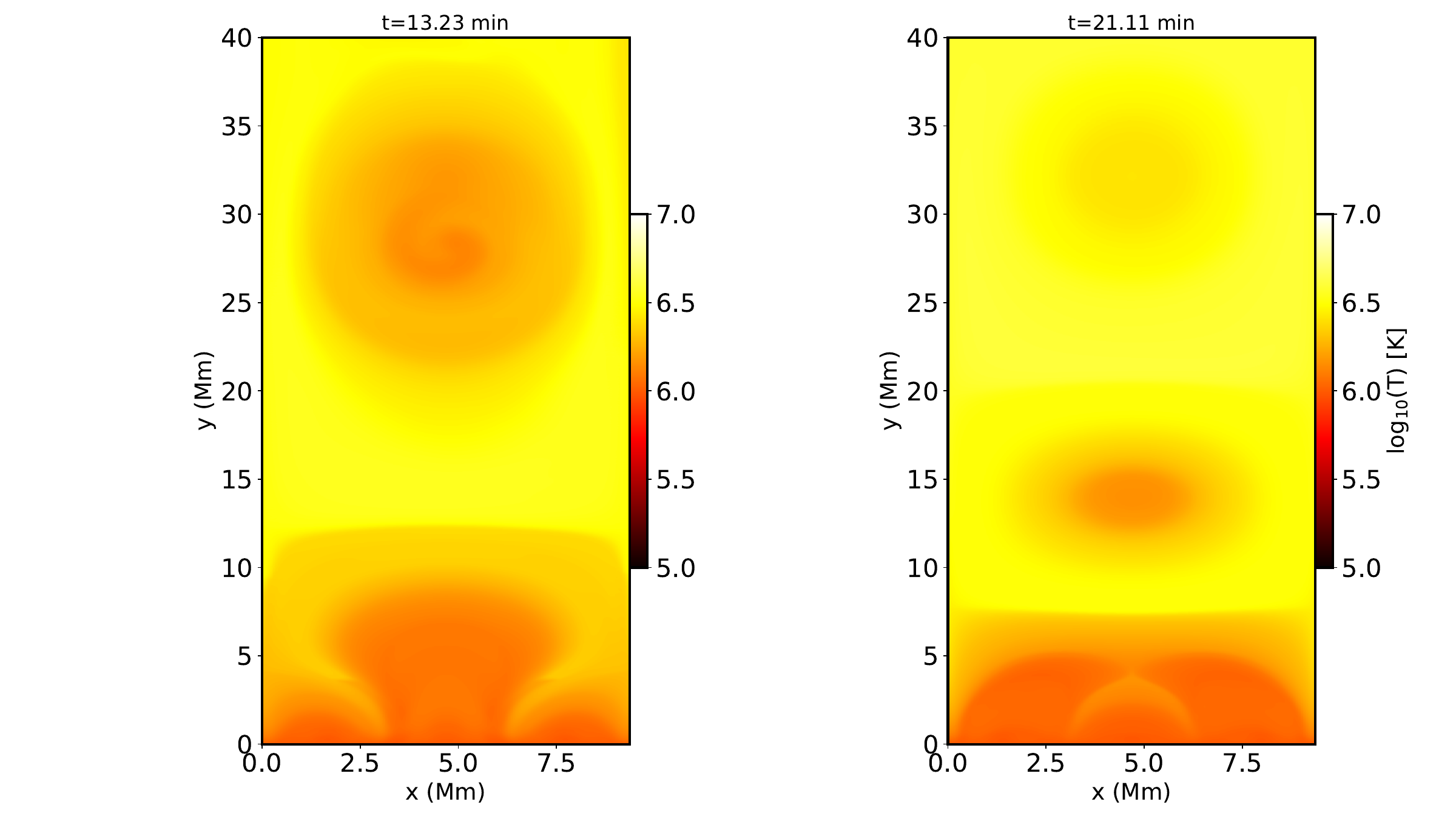}
\caption{Spatial distribution of plasma density (top panel), and temperature (bottom panel) for $t=13.23$ min, $21.11$ min, and $33.5$ min. The over-plotted solid curves represent the magnetic field lines projected in the $x-y$ plane (height and width aspect ratio of the figures are not in scale).} 
\label{fig:app}
\end{figure*}

\begin{figure*}[hbt!]
\centering
\includegraphics[width=0.8\textwidth]{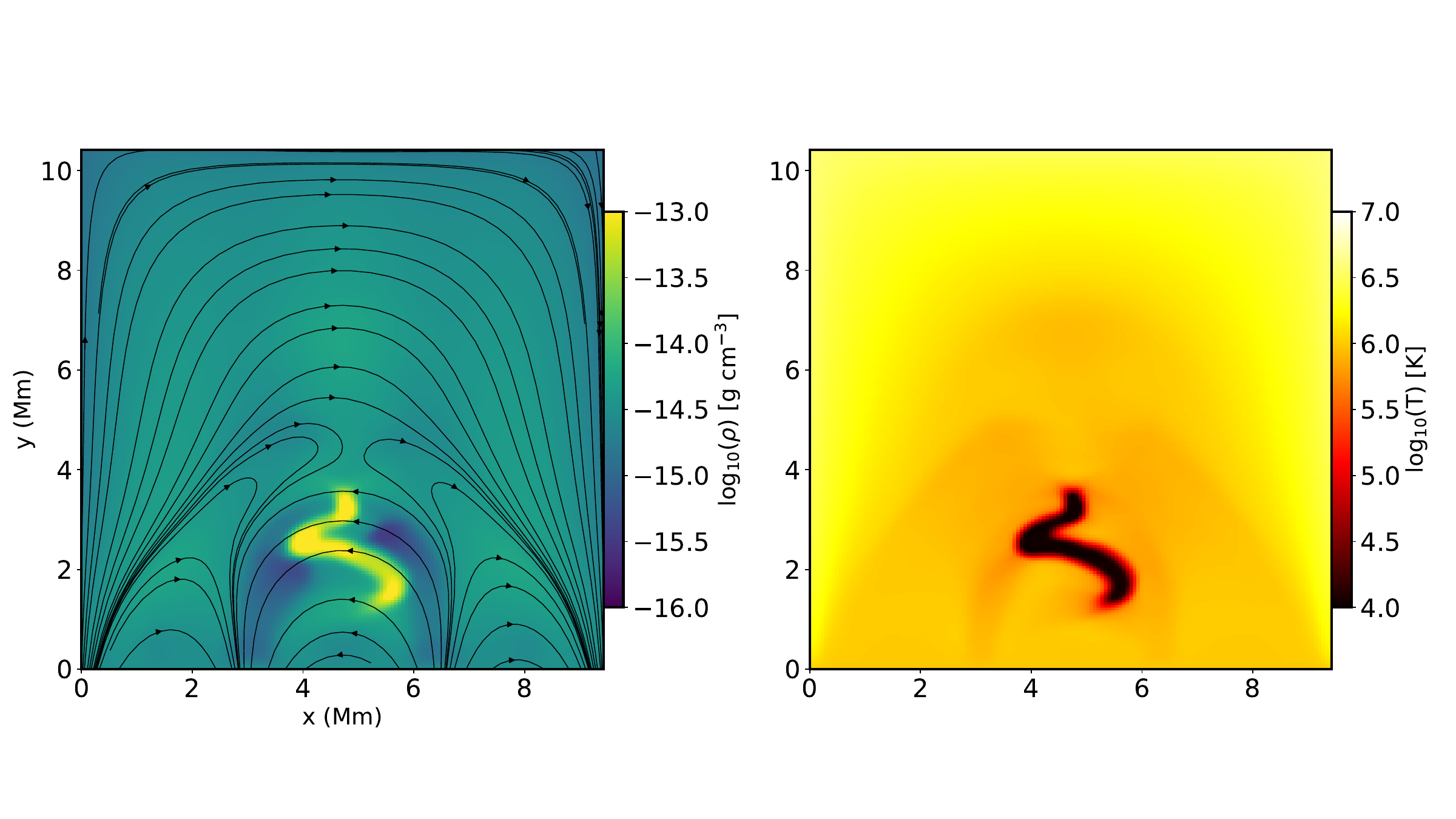}
\caption{Spatial distribution of density (left) and temperature (right) distribution at $t=35.78$ min, showing the splitting of the coronal rain blobs that forms at the loop top. The overplotted black curves represent the magnetic field lines projected in the $x-y$ plane.} 
\label{fig:app2}
\end{figure*}

We perform an additional simulation with a modified setup, where we use a three-arcade system by extending the lateral domain of our boundary from $x=0$ to $3\pi$ Mm (9.42 Mm), and the top boundary from y=0 to 40 Mm. We use the highest resolution of $65$ km and $54$ km along $x$ and $y$ directions respectively. Here, we modify the side boundary condition from periodic to the prescription according to \cite{2021A&A...646A.134J}, where the density and pressure are symmetric; $v_x$, $v_y$, $v_z$ are anti-symmetric, symmetric, and anti-symmetric respectively; $B_x$, $B_y$, and $B_z$ are anti-symmetric, symmetric, and anti-symmetric respectively at left and right boundaries. The top and bottom boundary conditions follow the same prescription as described in section \ref{sec:setup}. The density, temperature, and the magnetic field line evolution for the eruptions are shown in Figure \ref{fig:app}. We notice the formation of multiple flux ropes from the central arcade which expand laterally while eruption, and create horizontal field overlying the arcades. The magnetic field topology above the arcades changes drastically as successive flux ropes form and pass through. Finally, we notice the formation of coronal condensation at the central loop top at $t \approx 33$ min, which splits into different rain blobs and fall down, as shown in Figure \ref{fig:app2}. This time scale of condensation is earlier than the other setup (which is around 64 min). This difference in time scale appears due to the initial magnetic field structure with three arcades, modified boundary condition at the side, different spatial domain, and spatial resolution.

\end{appendix}

 

\end{document}